%% file: Ver-PRD-4.tex
\documentclass[preprint,floatfix,aps,prd,showpacs,nofootinbib,amsmath,amssymb,amsfonts,superscriptaddress]{revtex4-1}
\usepackage{graphicx}
\usepackage{color}
\usepackage{dsfont}
\usepackage[normalem]{ulem}
\usepackage{hyperref}
\usepackage{mathtools}
\usepackage{mathrsfs}
\usepackage{calrsfs}
\usepackage{comment}
\usepackage[utf8]{inputenc}
\usepackage{multirow} 
\usepackage{subfigure}
\usepackage{varioref}
\usepackage[active]{srcltx}

\expandafter\ifx\csname package@font\endcsname\relax\else
\expandafter\expandafter
\expandafter\usepackage
\expandafter\expandafter
\expandafter{\csname package@font\endcsname}%
\fi
\hypersetup{
	colorlinks=true,       
	linkcolor=blue,          
	citecolor=blue,        
}

\usepackage[section]{placeins}

\usepackage{fancybox}
\labelformat{figure}{Fig.~#1}

\begin{document}
	\title{Helical magnetic fields from Riemann coupling lead to 
baryogenesis}	
	\author{Ashu Kushwaha} 
	\email{ashu712@iitb.ac.in}
	\affiliation{Department of Physics, Indian Institute of Technology Bombay, Mumbai 400076, India}
	\author{S. Shankaranarayanan}
	\email{shanki@phy.iitb.ac.in}
	\affiliation{Department of Physics, Indian Institute of Technology Bombay, Mumbai 400076, India}
\begin{abstract}
The spectrum of energy density fluctuations, baryon asymmetry, and coherent large-scale magnetic fields are the three observables that provide crucial information on physics at very high energies. Inflation can only provide a mechanism to explain the density perturbations, and the origin of primordial magnetic fields and baryon asymmetry require physics beyond the standard models of cosmology and particle physics. In this work, we show that the mechanism that leads to primordial helical fields also leads to baryogenesis at the beginning of the radiation-dominated epoch. The model we consider here consists of mass dimension 6 operators that include Riemann coupling between gravity and electromagnetic field \emph{without} extending the Standard Model of particle physics. We explicitly show that the generation of primordial helical magnetic fields leads to baryogenesis. We further show that the model predicts the observed amount of baryon asymmetry of the Universe for a range of reheating temperatures consistent with the observations.
\end{abstract}
	\maketitle
\section{Introduction}
Understanding the physical processes in the very early Universe is a crucial ingredient for deciphering the physics at energies that we cannot currently probe in terrestrial experiments. While most observables have been washed away by the thermal bath of the pre-recombination era and do not have observational consequences, \emph{three observables} provide crucial information of the physics at high-energies. These are the spectrum of energy density fluctuations~\cite{Book-Kolb.Turner,Book-Mukhanov,Book-Padmanabhan-III,Book-Gorbunov.Rubakov}, excess of baryons over antibaryons (baryon asymmetry)~\cite{1999-Riotto.Trodden,2003-Dine.Kusenko-RevModPhy,2006-Cline-arXiv,2011-Riotto-JPCS,2007-Yoshimura-JPSJ,2015-Cui-ModPhyLetA,2020-Garbrecht-PPNP}, and coherent large-scale magnetic fields~\cite{2001-Grasso.etal-PhyRep,2002-Widrow-Rev.Mod.Phys.,2013-Durrer.Neronov-Arxiv,2016-Subramanian-Arxiv,2004-Giovannini-IJMPD,2020-Vachaspati-arXiv}. 

The inflationary paradigm provides an attractive mechanism to generate the primordial density perturbations that lead to anisotropies in the cosmic microwave background (CMB) and the formation of large-scale structures~\cite{Book-Kolb.Turner,Book-Mukhanov,Book-Padmanabhan-III,Book-Gorbunov.Rubakov}. During inflation, the early Universe underwent an accelerated expansion, stretching quantum fluctuations to super-horizon scale density perturbations. Besides providing a causal mechanism to density perturbations, inflation also solves the standard cosmological model's long-standing puzzles, such as the horizon, flatness, and monopole problems.

The predictions of inflation are in good agreement with the present-day observations of CMB anisotropies and polarization~\cite{2018-Planck}. However, within the standard electrodynamics, inflation cannot provide a mechanism to generate large-scale B fields. This is because in 4-dimensions electromagnetic field is conformally invariant. Since FRW models are conformally flat, the electromagnetic field vacuum in FRW is the same as the Minkowski space-time. Hence, the standard electromagnetic fields generate negligible magnetic fields. More importantly, even if the baryon asymmetry or cosmological magnetic fields existed before the epoch of inflation, these would have been diluted by a factor of $e^{-3 N}$, where $N$ is the number of e-foldings of inflation~\cite{2016-Fujita.Kamada-PRD,2019-Domcke.etal-JCAP,2014-Long.Sabancilar.Vachaspati-JCAP}. 

The present Universe is observed to contain essentially only matter and no antimatter, except for the rare antiparticles produced by cosmic rays. The asymmetry between baryons and antibaryons, referred to as Baryon Asymmetry of the Universe (BAU), can be expressed as~\cite{PartDataGroup,2018-Planck}
\begin{equation}
\label{def:etaObs}
\eta_B =\frac{n_{b}-n_{\bar{b}}}{n_{\gamma}}=\left\{\begin{array}{r}
{[5.8-6.6] \times 10^{-10}}~~\text{(from BBN)} \\
(6.09 \pm 0.06) \times 10^{-10}~~\text{(from CMB)}
\end{array}\right.
\end{equation}
where $n_{b}, n_{\bar{b}}, n_{\gamma}$ refer to the density of baryons, antibaryons and photons, respectively. 
Magnetic fields permeate the Universe. Coherent magnetic fields in spiral galaxies and clusters of galaxies have a magnitude of the order of $\mu$Gauss~\cite{2001-Grasso.etal-PhyRep,2013-Durrer.Neronov-Arxiv,2016-Subramanian-Arxiv,2002-Widrow-Rev.Mod.Phys.}. There is also indirect evidence of a lower limit of order $10^{-16}~$G for the magnetic field contained in the voids between galaxies and clusters of galaxies~\cite{2010-Neronov.Vovk-Sci}.  

The origin of primordial magnetic fields and baryon asymmetry of the Universe are still unresolved issues and require physics beyond the standard models of cosmology and particle physics. This leads to the following questions: As the Universe cooled, from the early Universe to today, what were the processes responsible for generating baryon asymmetry and large-scale magnetic fields? 
Are these processes cosmological or particle physics or both? 
Since both require physics beyond the standard model, there is a \emph{tantalizing possibility} that the same new physics can solve both. In this work, we consider such a possibility and show that the mechanism that leads to primordial helical magnetic fields also leads to baryogenesis at the beginning of the radiation-dominated epoch. Interestingly, our mechanism \emph{also requires} stretching of the primordial helical magnetic fields to super-horizon scales 
during inflation --- the same mechanism that leads to primordial density perturbations.

Before we discuss the model itself, it is necessary to understand the key ingredients to generate baryon-asymmetry and magnetic fields and why the same new physics can potentially solve both these problems~\cite{1967-Sakharov,1996-Davidson-PLB}. In 1967, Sakharov listed three necessary conditions for creating the BAU~\cite{1967-Sakharov,1999-Riotto.Trodden}: (1) baryon number violation, (2) charge ($C$) and charge parity ($CP$) violation, and (3) departure from thermal equilibrium. All three of the Sakharov conditions are satisfied in the Standard Model; however, the electroweak phase transition is not sufficiently strong in the first order~\cite{1999-Riotto.Trodden,2003-Dine.Kusenko-RevModPhy,2006-Cline-arXiv,2011-Riotto-JPCS}. The CP-violating effects are not sufficiently pronounced to account for as large a BAU as we observe. As a result, there must have been additional physics beyond the standard model to produce it. This physics could have been operating anywhere between the weak scale and the GUT scale. Corresponding to out-of-equilibrium conditions, the baryogenesis scenarios are divided into two categories: (a) by the universe expansion itself or (b) by fast phase transition and bubble nucleation. In particular, the latter concerns the electroweak baryogenesis schemes, while the former is typical for a GUT
type baryogenesis or leptogenesis~\cite{1999-Riotto.Trodden,2003-Dine.Kusenko-RevModPhy,2006-Cline-arXiv,2011-Riotto-JPCS}.

More than two decades ago, Davidson pointed out an interesting relation between the primordial magnetic field and Sakharov's conditions~\cite{1996-Davidson-PLB}. She argued that the presence of background magnetic fields in the early Universe could lead to the breaking of $C, CP, SO(3)$ symmetries and thermal equilibrium. Specifically, she argued that the presence of the magnetic fields leads to the following three conditions: (1) There should be some moderately out-of-thermal-equilibrium dynamics because in equilibrium, the photon distribution is thermal, and there are no particle currents to sustain a "long-range" field, (2) Since $\textbf{B}$ is odd under $C$ and $CP$, the presence of magnetic field will lead to $CP$ violation, (3) Since the magnetic field is a vector quantity, it chooses a particular direction hence breaks the isotropy (rotational invariance). Thus, Davidson provided a possible link between the presence of magnetic fields to the conditions required for baryogenesis~\cite{1996-Davidson-PLB}.

Davidson's conditions are necessary \emph{but not} sufficient. One key missing ingredient, as we show, is the requirement of \emph{primordial helical magnetic fields} (details in Sec. \ref{sec:Baryo-magnetic}). Primordial helical magnetic fields are generated by the terms that break conformal invariance and parity symmetry~\cite{2001-Vachaspati-PRL,2003-Caprini.etal-PRD,2005-Campanelli-Giannotti-PRD,2018-Sharma.Subramanian.Seshadri.PRD,2009-Caprini.Durrer.Fenu-JCAP,2009-Campanelli-IJMPD,2019-Shtanov-Ukr.PJ,2020-Kushwaha.Shankaranarayanan-PRD}.
If we could measure them, primordial helical magnetic fields provide evidence of CP violation in the early Universe. Interestingly, the presence of primordial helical fields leads to non-zero Chern-Simons number~\cite{2016-Fujita.Kamada-PRD,2015-Anber.Sabancilar-PRD,2016-Kamada.Long-PRD} and, eventually, the change in the Fermion number. 

Recently, the current authors constructed a simple model of inflationary magnetogenesis that couples the electromagnetic fields with the Riemann tensor~\cite{2020-Kushwaha.Shankaranarayanan-PRD}. We showed that this model leads to a primordial helical magnetic field where one helical mode is enhanced while the other mode is suppressed. The model has two key advantages over other models~\cite{2005-Campanelli-Giannotti-PRD,2018-Sharma.Subramanian.Seshadri.PRD,2009-Caprini.Durrer.Fenu-JCAP,2009-Campanelli-IJMPD,2019-Shtanov-Ukr.PJ}: First, it does not require the coupling of the electromagnetic field with any scalar field. Hence, unlike Ratra model~\cite{1991-Ratra-Apj.Lett,2019-Shakeri.etal-PRD,2009-Demozzi.etal-JCAP}, there is no strong-coupling problem caused by the extra degrees of freedom. 
Second, the model is free from backreaction for generic slow-roll inflation models~\cite{2020-Kushwaha.Shankaranarayanan-PRD}. In Ref.~\cite{2018-Sharma.Subramanian.Seshadri.PRD}, authors have shown the strong-coupling problem in Ratra model~\cite{1991-Ratra-Apj.Lett} can be avoided by choosing a particular coupling function.

In that work, the current authors used the general effective field theory of gravity coupled to the Standard Model of particle physics framework to obtain leading order gravity terms that couple to the standard model Bosons~\cite{2019-Ruhdorfer.etal-JHEP}. As we have done in the previous work, we limit to mass dimension 6-operators coupling to the gauge field Lagrangian, specifically, to the electromagnetic field.

In this work also, we limit to mass dimension 6-operators coupling to the gauge field, specifically, to the electromagnetic field.
We show that the generation of primordial helical magnetic fields from the above model leads to baryogenesis. Since the model produces helical fields over large length scales, we show that the Chern-Simons (CS) number density is non-zero (details in Sec. \ref{sec:Baryo-magnetic}). Considering that the model generates primordial helical modes at all length scales, we focus on the last ten e-foldings of inflation. This is because the modes that leave the Hubble radius during the last 10 e-foldings of inflation will reenter the Universe after reheating; these primordial helical modes will lead to baryogenesis just at the beginning of the radiation-dominated epoch.  Furthermore, we show that the BAU is independent of inflation models and depends \emph{only on} the energy scale at the exit of inflation and reheating temperature.

In Sec. (\ref{sec:Baryo-magnetic}), we discuss the relation between primordial helical magnetic fields and baryogenesis, in particular, the chiral anomaly in the presence of the magnetic field, and obtain the expression for Chern-Simon number density. In Sec. (\ref{sec:Model}), we discuss the generation of primordial helical modes and show that primordial helical modes lead to a non-zero CS number density. Then we evaluate the baryon asymmetry parameter in Sec.(\ref{sec:baryon_asymm}). Sec. \eqref{sec:conc} contains the implications of the results. Appendices contain the details of the calculations.

In this work, we use $(+,-,-,-)$ signature for the 4-D space-time metric. Greek alphabets denote the 4-dimensional space-time coordinates, and Latin alphabets denote the 3-dimensional spatial coordinates.  \emph{A prime} stands for a derivative with respect to conformal time $(\eta)$ and \emph{subscript} $,i$ denotes a derivative w.r.t spatial coordinates. We use the Heaviside-Lorentz units such that $c = k_B = \epsilon_0 = \mu_0 = 1$. The reduced Planck mass is denoted by $M_{\rm P} = (8 \pi G)^{-1/2}$.

%
\section{Conditions on baryogenesis in the presence of primordial magnetic field}
\label{sec:Baryo-magnetic}

As we mentioned in the introduction, Davidson's conditions are necessary but not sufficient. One key missing ingredient is the requirement of \emph{primordial helical magnetic fields}. In this section, we briefly discuss this. 

In the very early Universe, just after the exit of inflation, the energy scale of the Universe was close to $10^{14}~{\rm GeV}$. All particles, including Fermions, are highly relativistic and can be treated as massless. 
Although the massless Dirac equation is invariant under chiral transformations in the classical theory, the chiral symmetry is broken due to quantum mechanical effects in the presence of the external electromagnetic fields. This phenomenon, known as the quantum axial anomaly, affects the transport properties of the chiral medium, leading to experimentally accessible signatures such as the chiral magnetic effect~\cite{2008-Fukushima.etal-PRD} and the 
chiral separation effect~\cite{2005-Metlitski.Zhitnitsky-PRD}.

In the early Universe, the generation of the non-zero primordial helical magnetic fields leads to a chiral anomaly resulting from the imbalance between left and right-handed fermions. 
In  the presence of an electromagnetic field in curved space-time, the chiral anomaly is given by the following equation~\cite{2009-Parker.Toms-Book,2014-Barrie.Kobakhidze-JHEP}:
\begin{align}\label{eq:chiralAnomaly}
 \nabla_{\mu}J_A^{\mu}  = -\frac{1}{384 \pi^2} \epsilon^{\mu\nu\rho\sigma}  R_{\mu\nu\alpha\beta} R^{\alpha\beta}\,_{\rho\sigma} + \frac{e^2}{16 \pi^2} \epsilon^{\mu\nu\alpha\beta} F_{\mu\nu} F_{\alpha\beta}  
\end{align}
where $J^{\mu}_A$ is the chiral current, $R_{\rho\sigma}\,^{\alpha\beta}$ is the Riemann tensor and  $A_{\mu}$ is the four-vector potential of the electromagnetic field, $F_{\mu\nu} = \nabla_{\mu}A_{\nu} - \nabla_{\nu}A_{\mu} $. $\epsilon^{\mu\nu\rho\sigma} = \frac{1}{\sqrt{-g}}\, \eta^{\mu\nu\rho\sigma}$ is a fully antisymmetric tensor, $\eta^{\mu\nu\rho\sigma}$ is Levi-Civita symbol whose values are $\pm1$ and we set $\eta^{0123} = 1 = - \eta_{0123}$. It is easy to see from the above equation that the
anomaly contribution from the electromagnetic field and the gravity act independently and, for most parts, can be treated independently.

In the case of flat FRW background in conformal time ($\eta$):
\begin{align}\label{eq:FRW}
ds^2 = a^2(\eta) \,(d\eta^2 - \delta_{ij} dx^i dx^j)
\end{align}
the contribution of the first term in the RHS of Eq.~(\ref{eq:chiralAnomaly}) vanishes, i. e.,
\begin{align}
\epsilon^{\mu\nu\rho\sigma} R_{\mu\nu\alpha\beta} R^{\alpha\beta}\,_{\rho\sigma} = 0 \, .
\end{align}
It can be shown that even at the first-order, the gravitational contribution vanishes, and the non-zero contribution arises only at second order~\cite{2006-Alexander.Peskin.Jabbari-PRL}. Due to the presence of the antisymmetric tensor, the gravitational fluctuations lead to gravitational birefringence and can lead to net chiral current. 

In the flat FRW background, the second term in the RHS of Eq.(\ref{eq:chiralAnomaly}) is given by:
\begin{align}
\frac{e^2}{16 \pi^2} \epsilon^{\mu\nu\alpha\beta} F_{\mu\nu} F_{\alpha\beta} = \frac{e^2}{4 a^4} \epsilon_{ijk} \partial_j A_k \, \partial_0 A_i  \, .
\end{align}
In the presence of the magnetic field, this term is non-zero and hence leads to 
a net chiral current. 
Thus, if we consider only up to the first-order in perturbations, 
\emph{only} the second term in the RHS of  Eq.~(\ref{eq:chiralAnomaly})  contributes and the chiral anomaly equation reduces to:
\begin{align}\label{eq:chiralAnomaly_FF}
\partial_{\mu}\left(  \sqrt{-g} J_A^{\mu} \right) =  \frac{e^2}{16 \pi^2} \eta^{\mu\nu\alpha\beta} F_{\mu\nu} F_{\alpha\beta}  \, ,
\end{align}
where we have used 
\[
\nabla_{\mu}J^{\mu}_A = \frac{1}{\sqrt{-g}} \partial_{\mu} \left( \sqrt{-g} J^{\mu}_A \right), \quad \epsilon^{\mu\nu\alpha\beta} = \frac{1}{\sqrt{-g}} \eta^{\mu\nu\alpha\beta} \, .
\]

Note that during inflation, LHS in Eq. (\ref{eq:chiralAnomaly_FF}) is zero, 
and due to the exponential expansion, standard model particles are diluted. However, if we can generate non-zero primordial helical fields during inflation, then 
these non-zero primordial helical fields can lead to chiral current at the radiation-dominated epoch (or during reheating when the standard model particles are created). To see this, we rewrite Eq.~(\ref{eq:chiralAnomaly_FF})
using $\eta^{\mu\nu\alpha\beta} F_{\mu\nu} F_{\alpha\beta} = 4 \partial_{\mu} \left( \eta^{\mu\nu\alpha\beta} A_{\nu} \partial_{\alpha} A_{\beta}  \right)$, i.e.,
\begin{align}\label{eq:chiral_topo_current}
 \partial_{\mu} \left( \sqrt{-g} J^{\mu}_A \right)  =  \frac{e^2}{4 \pi^2} \partial_{\mu} \left(  \eta^{\mu\nu\alpha\beta} A_{\nu} \partial_{\alpha} A_{\beta}  \right) =  \frac{e^2}{4 \pi^2} \partial_{\mu} \left( \sqrt{-g} K^{\mu} \right) 
\end{align}
where 
\[
K^{\mu} =  \frac{\eta^{\mu\nu\alpha\beta} }{\sqrt{-g} } A_{\nu} \partial_{\alpha} A_{\beta} 
\]
is the topological current. For FRW background, the components are given by
\begin{align}
K^0 =  a^{-4}(\eta) \, \epsilon_{ijk} A_{i} \partial_j A_k  \qquad \text{and} \qquad K^i = a^{-4}(\eta) \, \epsilon_{ijk} A_{j} \partial_0 A_k.
\end{align}
Solving Eq.~(\ref{eq:chiral_topo_current}), we get, 
\[
J^{\mu}_A   =  \frac{e^2}{4 \pi^2} K^{\mu}  \, .
\]
Thus, the net baryon number density, $n_B = n_b - n_{\bar{b}} = a(\eta) \langle 0 | J^0_A | 0 \rangle$ is related to Chern-Simon number density $n_{CS} = \langle 0 | K^0 | 0 \rangle$ as~\cite{2014-Barrie.Kobakhidze-JHEP},
\begin{align}\label{eq:n_B-n_CS-definition}
n_B \equiv \frac{e^2}{4\pi^2} a(\eta) n_{CS}.
\end{align}
Note that $n_{CS} = 0$ at the start of inflation, and due to the absence of standard model particles $n_B = 0$ during inflation. Using the expression for $K^0$, we can write the Chern-Simon number density as
\begin{align}\label{eq:n_cs-relation}
n_{CS} = \frac{1}{a^4} \epsilon_{i j k} \langle 0 | A_i \, \partial_j A_k | 0 \rangle = \frac{1}{a^4}\int_{\mu}^{\Lambda} \frac{dk}{k} \frac{k^4}{2\pi^2} \left(  | A_+ |^2 - |A_-|^2  \right) \, ,
\end{align}
where $\Lambda$, and $\mu$ set the possible energy range (or epoch) during which baryon asymmetry is generated after inflation, and $A_{\pm}$ refer to the positive and negative helicity modes of the electromagnetic field. The above expression is key in illuminating a useful relation between primordial helical magnetic fields generated during inflation and baryogenesis: 
First, we see that the contribution to $n_{CS}$ is from all the modes that reenter the horizon at the beginning of the radiation-dominated epoch. Thus, the value of $n_{CS}$ depends on the upper cut-off $\Lambda$. 
Second, the expression corresponds to the total Chern-Simons number density generated from the modes 
in the energy range $[\mu, \Lambda]$ ---  when these helical modes re-enter during the radiation-dominated epoch. 
The helicity modes $A_+ $ and $ A_-$ are generated during inflation, and ${a^{-4}(\eta)}$ is the dilution due to the expansion of the Universe during this epoch. 
Finally, $n_{CS}$ vanishes if the primordial magnetic fields are non-helical, i. e. $|A_+ | = |A_-|$. Hence, as mentioned at the beginning of this section, the generation of non-helical magnetic fields will not lead to baryogenesis. Thus, the key missing ingredient of Davidson's argument is the requirement of primordial helical magnetic fields. 

In the following two sections, we explicitly evaluate the Chern-Simons number for our model and show that it is not sensitive to inflationary and reheating dynamics. 
\section{The model and the primordial helical fields}
\label{sec:Model}
We consider the following action \cite{2020-Kushwaha.Shankaranarayanan-PRD} :
\begin{align}\label{eq:action}
S  = S_{\rm{Grav}} + S_{\phi} + S_{\rm{EM}} + S_{\rm CB}
\end{align}
where $ S_{\rm{Grav}}$ is the Einstein-Hilbert action
\begin{align}\label{eq:EH-action}
S_{\rm Grav} = -\frac{M_{\rm P}^2}{2}\int d^4x \sqrt{-g} \, R \, ,
\end{align}
and $ S_{\phi} $ is the action for the minimally coupled, self-interacting canonical scalar field:
\begin{align}\label{eq:inflation-action}
S_{\phi} = \int d^4x \sqrt{-g} \left[  \frac{1}{2} \partial_{\mu}\phi \partial^{\mu}\phi -  V(\phi) \right].
\end{align}
$S_{\rm{EM}}, S_{\rm CB}$ refer to the standard electromagnetic (EM)  and conformal breaking part of the electromagnetic terms, respectively, which are given by:
\begin{align}\label{eq:S_EM}
 S_{\rm{EM}} &= -\frac{1}{4} \int d^4x \, \sqrt{-g} \, F_{\mu\nu} F^{\mu\nu}, \hspace{0.5cm}\\  
 \label{eq:S_h}
 S_{\rm{CB}} &= - \frac{1}{M^2} \,\int d^4x \, \sqrt{-g} \, R_{\rho\sigma}\,^{\alpha\beta} F_{\alpha\beta} \, \tilde{F}^{\rho\sigma} = - \frac{1}{M^2} \,\int d^4x \, \sqrt{-g} \, \tilde{R}^{\mu\nu\alpha\beta} F_{\alpha\beta} \, F_{\mu\nu} \, ,
 \end{align}
where $\tilde{R}^{\mu\nu\alpha\beta} = \frac{1}{2}\epsilon^{\mu\nu\rho\sigma} R_{\rho\sigma}\,^{\alpha\beta}$ is the dual of Riemann tensor and $\tilde{F}^{\rho\sigma} = \frac{1}{2} \epsilon^{\mu\nu\rho\sigma}F_{\mu\nu} $ is the dual of $F_{\mu\nu}$.  The standard electromagnetic action $S_{\rm{EM}}$ is conformally invariant; however, the presence of Riemann curvature in $S_{\rm CB}$ breaks the conformal invariance. $M$ is the energy scale, which sets the scale for the breaking of conformal invariance. Note that the signs of $S_{\rm{EM}}$ and $S_{\rm{CB}} $ are chosen with respect to the positive electromagnetic energy density.

In Ref. \cite{2019-Ruhdorfer.etal-JHEP}, the authors systematically showed that the first gravity operators appear at mass dimension 6 in the series expansion of the coupling between gravity and the standard model of particle physics. These operators only couple to the standard model Bosons. They also showed that (i) no new gravity operators appear at mass dimension 7, (ii) in mass dimension 8, the standard model Fermions
appear, and (iii) coupling between the scalar (Higgs) field and the standard model gauge Bosons appear only at mass dimension 8. Since mass dimension 8 operators are highly suppressed, like in Ref. \cite{2020-Kushwaha.Shankaranarayanan-PRD}, we limit ourselves to mass dimension 6  operators. Due to Riemann coupling, $M$ appears as a time-dependent coupling in the FRW background i.e., $1/M_{\rm eff} \sim H/M$.
At the current epoch where 
$H_0 \approx 10^{-42} \rm{GeV}$ and assuming the parameter $M \approx 10^{17} \rm{GeV}$, we obtain ${H_0}/{M} \sim 10^{-59}$. Therefore, the coupling (Riemann tensor) is tiny and the non-minimal coupling term in the electromagnetic action will have significant contribution only in the early universe.
We also would like to point that the coupling term  
($S_{\rm CB}$) is tiny near the Schwarzschild radius of a solar mass black-hole  
(for details, see appendix \ref{app:blackhole}).

We assume that the scalar field ($\phi$) dominates the energy density in the during inflation and leads to $60 \, - \, 70$ e-foldings of inflation with $H_{\rm Inf} \sim 10^{14} {\rm GeV}$. Specifically, we consider power-law inflation in which the scale factor (in conformal time) is~\cite{2004-Shankaranarayanan.Sriramkumar-PRD}:
 \begin{align}\label{eq:powerLaw}
a(\eta) =  \left( - \frac{\eta}{\eta_0} \right)^{(\beta+1)}
\end{align}
where, the constant $\eta_0$ denotes the scale of inflation and $\beta \leq -2$. $\beta = -2$ corresponds to exact de Sitter.  During inflation, $\eta \in (-\infty, 0)$. For slow-roll inflation $ \beta \approx -2-\epsilon$ and $ \mathcal{H} \equiv a^{\prime}/{a} \approx - (1 + \epsilon)/{\eta}$, where 
$\mathcal{H}$ is the Hubble parameter in conformal time and 
$\epsilon $ is the slow roll parameter. For our discussion below, we also assume that $10^{-3}  \leq (H_{\rm Inf}/M) \leq 1$~\cite{2018-Nakonieczny-JHEP,2016-Goon.Hinterbichler-JHEP,2016-Goon-JHEP,2013-Balakin.etal-CQG}.

Equation of motion of the gauge field can be obtained by varying the action (\ref{eq:action}) with respect to $A^{\mu}$. In the Coulomb gauge ($A^{0} = 0, \partial_iA^i = 0$), we have:
\begin{align}\label{eq:equation_of_motion}
A_i^{\prime\prime} + \frac{4 \, \epsilon_{i j l}}{M^2} \, \left( \frac{a^{\prime\prime\prime}}{a^3} - 3\frac{a^{\prime\prime} a^{\prime} }{a^4} \right) \partial_j A_l 
- \partial_j \partial_j A_i = 0
\end{align}
where $\epsilon_{i j l}$ is the Levi-Civita symbol in the 3-D Euclidean space. The above equation is different from other models in the literature and leads to distinct evolution of the magnetic field fluctuations in comparison to non-minimally coupled scalar field models~\cite{2020-Kushwaha.Shankaranarayanan-PRD}. In the helicity basis, the above equation reduces to (see appendix \ref{app:helicity_basis}):
\begin{align}\label{eq:eom_helicity}
A_h^{\prime\prime} + \left[  k^2 - \frac{4kh}{M^2} \, 
\left( \frac{a^{\prime\prime\prime}}{a^3} - 3\frac{a^{\prime\prime} a^{\prime} }{a^4} \right)  \right] A_h= 0 \, .
\end{align}
For the two helicity states ($h = \pm$), the above expression leads to two different evolution equations [cf. Eqs.~(\ref{eq:sup_mode_h+}, \ref{eq:sup_mode_h-})]. From Eq. \eqref{eq:n_cs-relation} we see that to obtain appreciable value of Chern-Simons number ($n_{CS}$), the difference between the two helicity states should be non-zero, and it is maximum 
if one helicity mode is enhanced compared to other. 

In our previous work \cite{2020-Kushwaha.Shankaranarayanan-PRD}, we showed that for a range of parameters of interest, negative helicity mode decays while the positive helicity mode is enhanced. Hence, negative helicity mode ($A_-$) will have negligible contribution and can be set to zero, i. e., 
$|A_-| = 0$. Using the series expansion of the Bessel functions, in the leading order, the positive helicity mode takes the following form
(\ref{eq:sup_mode_h+}): 
%
\begin{align}\label{eq:A+Series}
A_+(\tau,k) &= C \, k^{\frac{1}{4\alpha}} 
  - C_2 \frac{\mathcal{F}^{-1} }{\pi} 
  \Gamma \left( \frac{1}{2\alpha}  \right) \,k^{-\frac{1}{4\alpha}} \tau^{ - \frac{1}{\alpha} }  
  %
\end{align}
where,
\begin{align}\label{eq:C_F}
| C | \approx \varsigma^{-1} |C_2| \approx 
\frac{M^{3/2} \eta_0}{\sqrt[4]{\eta_{end} 10^{45} GeV^3}},~~  
\mathcal{F} \approx |\varsigma|^{-1} \approx \sqrt{M^2 \eta_0}, ~~
%
%
\alpha = -\frac{1}{2} -\epsilon
\end{align}
For details, see Appendix \eqref{app:Helical}. 

Our model generates primordial magnetic fields through the non-minimal coupling of the electromagnetic field. The model requires inflation. Inflation generates density perturbations at all scales and provides a causal mechanism to generate the structure formation. Similarly, our model generates magnetic fields at all length scales, including the current Horizon radius~\cite{2001-Grasso.etal-PhyRep,2013-Durrer.Neronov-Arxiv,2016-Subramanian-Arxiv,2002-Widrow-Rev.Mod.Phys.,2004-Giovannini-IJMPD}. This has to be contrasted from the models where the magnetic field is generated during recombination. In these models, the coherence scale of the generated fields cannot exceed the size of the horizon radius at that time.

In Appendix \ref{app:Helical}, we have plotted the power spectrum of the present-day helical magnetic field ($B_0$) as a function of $k$. Assuming $M = 10^{17} GeV$, our model predicts the  primordial helical magnetic fields of strength $10^{-20} \rm{G}$ on Gpc  scales at the current epoch. From \ref{fig:PowerSpectrum} we can see that our model predicts the present-day helical magnetic field of strength $10^{-15} G$ on Mpc scales. The primordial fields generated from our model are within the upper bounds on the strength of the seed magnetic fields needed to explain the current galactic magnetic fields~\cite{2010-Kahniashvili.Ratra.etal-PRD}. These primordial fields are amplified by the dynamo mechanism and can lead to the observed magnetic fields; hence our model requires the dynamo mechanism.

%
%
%
\section{Baryon Asymmetry of the Universe}
\label{sec:baryon_asymm}
In this section, we compute the baryon asymmetry parameter due to the primordial helical magnetic fields. Specifically, we compute it for the maximum helicity modes --- one mode is enhanced compared to the other. Substituting Eq~(\ref{eq:A+Series}) in Eq.~(\ref{eq:n_cs-relation}), we obtain
\begin{align}\label{eq:n_CS-integral}
n_{CS} = \frac{1}{2\pi^2 \, a^4(\eta)}\int^{\Lambda}_{\mu} dk  \left( \left| C \right|^2 \, k^{3 + \frac{1}{2\alpha}} 
  + \left| C_2 \frac{\mathcal{F}^{-1} }{\pi} 
  \Gamma \left( \frac{1}{2\alpha}  \right) \right|^2\,k^{3 -\frac{1}{2\alpha}} \tau^{ - \frac{2}{\alpha} }    \right).
\end{align}
Integrating the above expression, we get
\begin{align}\label{eq:n_CS}
n_{CS} =  \frac{1}{2\pi^2 \, a^4(\eta)}\left[  \left.  \left| C \right|^2 \,  \frac{ k^{4 + \frac{1}{2\alpha}} }{ 4 + \frac{1}{2\alpha}} \right|^{\Lambda}_{\mu}
  + \left. \left| C_2 \frac{\mathcal{F}^{-1} }{\pi} 
  \Gamma \left( \frac{1}{2\alpha}  \right) \right|^2 \,\frac{ k^{4 -  \frac{1}{2\alpha}} }{ 4 - \frac{1}{2\alpha}} \tau^{ - \frac{2}{\alpha} } \right|^{\Lambda}_{\mu} \,\,    \right].
\end{align}
We want to make the following remarks regarding the above expression: First, the BAU is generated similarly to the inflationary mechanism of the generation of density perturbation. During inflation, the primordial helical magnetic field fluctuations are stretched exponentially and exit the horizon. The modes that reenter during the radiation-dominated epoch are responsible for the generation of baryon asymmetry.  Second, the generation of baryon asymmetry does not strongly depend on the reheating dynamics since only the modes that 
reenter the Hubble radius during the radiation-dominated epoch are relevant.  

Assuming a de-Sitter (or approximately de-Sitter) Universe, from Eq. \eqref{eq:C_F}, 
we have $\tau^{-\frac{2}{\alpha}} = a^{-2}(\eta)$. Substituting this in the 
Eq. (\ref{eq:n_CS}), we see that the 
the second term in the RHS decays faster compared to the first term by $a^{-2}(\eta)$. Hence, we can neglect the second term. Substituting the 
resulting form of $n_{CS}$ in Eq.~(\ref{eq:n_B-n_CS-definition}) leads to:
\begin{align}\label{eq:n_B}
n_{B} = \frac{e^2}{4\pi^2} \frac{1}{2\pi^2 \, a^3(\eta)}  \left.  \left| C \right|^2 \,  \frac{ k^{4 + \frac{1}{2\alpha}} }{ 4 + \frac{1}{2\alpha}} \right|^{\Lambda}_{\mu}.
\end{align}
To obtain the ranges of $\Lambda$ and $\mu$,  we need to know the modes exited during inflation. For the density perturbations, the largest scales observed in the CMB are produced around 40 - 60 e-foldings before the end of inflation~\cite{2006-Bassett.Tsujikawa.Wands-RevModPhys}.
This is because the adiabatic quantum fluctuations responsible for the density perturbations reenter the Hubble radius around $z \sim 1500$. Hence, in Ref. \cite{2020-Kushwaha.Shankaranarayanan-PRD}, the current authors only looked at primordial helical fields generated around 40 - 60 e-foldings before the end of inflation. However, in this case, we will concentrate on the primordial helical fields that renter the horizon very early (at the beginning of the radiation-dominated epoch) to generate the required BAU. This means that the modes that left the horizon around the last 5 to 10 e-foldings of inflation are only relevant. Since these modes have already left the Hubble radius during inflation, the reheating dynamics do not alter these primordial helical modes. Hence, the model is insensitive to the reheating dynamics.  

Our focus now shifts to explicitly evaluating BAU for our model.  First step is to evaluate the 
dilution factor $a^{-3}$ in Eq. \eqref{eq:n_B}. To do this, we define 
$a_{\Lambda}$ (and $a_{\mu}$) as the scale factor at the time when the maximal helicity mode with energy $\Lambda$ (and $\mu$) left the Hubble radius during inflation. Assuming an instant reheating, and following the calculations given in Appendix \eqref{app:Calculations}, we have 
$a_{\mu} = 10^6 a_{\Lambda}$. Taking into account that these modes 
exited the Hubble radius during inflation in the last 5 e-foldings, the 
the dilution factor [prefactor in Eq. \eqref{eq:n_B}]  
becomes $a^{-3} \sim 10^{-24}$. 

The second step is to obtain the constant $C$. 
As discussed in previous section, for slow-roll inflation, $|C|$ is given by 
Eq. \eqref{eq:C_F}. Thus, Eq. \eqref{eq:n_B} reduces to:
\begin{align}\label{eq:n_B-Lambda}
n_{B} \approx  \quad \frac{10^{-24} \cdot |C|^2 \cdot e^2}{24\pi^4}  \left( \Lambda^3 - \mu^3 \right) \, .
\end{align}
Third step is to compare the theoretically derived quantity ($n_B$) 
with observations Eq.~\eqref{def:etaObs}. However, $n_{\gamma}$ is not constant in the early Universe (since the photon chemical potential is zero) and is approximately constant only after the last scattering surface. 
Since entropy density per comoving volume is conserved, the quantity $n_B/s$ is better suited for theoretical calculations~\cite{Book-Kolb.Turner}.
Assuming that there was no significant entropy production after reheating phase, entropy density in the radiation-dominated epoch is: 
\begin{equation}
s \simeq  \frac{2\pi^2}{45} g \,  T^3_{\rm{RH}} \, ,
\label{def:entdens}
\end{equation}
where $T_{\rm{RH}}$ is the reheating temperature and the effective relativistic degrees of freedom $g \sim 100$ at reheating. From Eqs. (\ref{eq:n_B-Lambda}, \ref{def:entdens}), we can define the following dimensionless BAU parameter: 
\begin{align}\label{eq:baryon_Asym}
\eta_B = \frac{n_B}{s} \approx 10^{-24} 
\frac{|C|^2 \cdot e^2}{24\pi^4 }  \left( \Lambda^3 - \mu^3 \right)
\frac{45}{2\pi^2 g T_{RH}^3}  \approx 10^{-29} |C|^2
\frac{\Lambda^3 }{  T^3_{\rm{RH}}  }
\end{align}
where in the last expression we have neglected $\mu^3$ i.e., $ \Lambda^3 - \mu^3 \approx \Lambda^3$. Appendix \eqref{app:Calculations} contains
plots for different values of $\Lambda$ and $\mu$. From these plots, we infer that the results do not strongly depend on the exact value of $\mu$. 

Finally, substituting the value of $|C|^2$ (from Eq. \eqref{eq:C_F} and 
using the values in Appendix \ref{app:Helical}) in Eq.~\eqref{eq:baryon_Asym}, we obtain:
\begin{align}\label{eq:baryon_Asym-M}
\eta_B  \approx  \frac{ 10^{-29} \cdot \eta_0^2 }{ \sqrt{\eta_{end} \cdot 10^{45} GeV^3} } \frac{M^3 \Lambda^3 }{  T^3_{\rm{RH}}  }  \approx 
10^{-2} \left( \frac{M}{M_P} \right)^3  
\left( \frac{\Lambda}{T_{ \rm{RH}}} \right)^3 
\end{align}
This is one of the crucial expressions in this work regarding which we would like to stress the following: First, the BAU parameter depends on three quantities --- $M$ (the conformal invariance breaking scale), $T_{\rm{RH}}$ (reheating temperature scale) and $\Lambda$ (the largest helical mode that catalyses
baryogenesis). 
Second, the BAU parameter is inversely proportional to the reheating temperature. This behavior is different from the results of Ref. \cite{2006-Alexander.Peskin.Jabbari-PRL,2014-Barrie.Kobakhidze-JHEP,2014-Long.Sabancilar.Vachaspati-JCAP,2016-Fujita.Kamada-PRD}. In some of these models, BAU is linearly dependent on the reheating temperature. The difference in the relationship is because the detailed reheating dynamics is not required, only the information about the entropy production is required in our model. In other models, the exact detailed reheating dynamics is required, which is avoided in our approach. 
Third, the BAU parameter is linearly proportional to $M$ and $\Lambda$. For smaller $M$, the contribution of the conformal breaking term \eqref{eq:S_h} will be much larger, and hence, more primordial helical fields are produced during inflation. However, for the same reheating temperature, $\Lambda$ has to be larger to produce the same amount of BAU. 
Fourth, to get a better understanding of the dependence of BAU on various parameters, we use the following parametrization:
\begin{align}\label{eq:parametrization}
\eta_B = n \times 10^{-10}, \quad M = m \times 10^{14} GeV,  
\quad \Lambda = \delta \times 10^{12} GeV, \quad 
T_{RH} = \gamma \times 10^{12} GeV 
\end{align}
where $n, m, \delta, \gamma$ are dimensionless parameters. The maximum reheating 
corresponds to the inflation scale~\cite{2006-Bassett.Tsujikawa.Wands-RevModPhys}. With supersymmetry, the 
requirement that not too many gravitinos are produced 
after inflation provides a stringent constraint on the reheating temperature,
$T_{\rm RH} \sim 10^{10} - 10^{11}~$GeV~\cite{1984-Ellis.Kim.Nanopoulos-PLB,1999-Benakli.Davidson-PRD}. Hence, we consider the range of $\gamma$ 
to be  $\{ 10^{-2}, 1000  \} $. Since the value of $M$ should be between the GUT and Planck scale, we consider the range of $m$ to be $\{ 1, 1000  \}$.  
We assume that the modes that reenter during radiation epoch is around $10^{12}~$GeV. Hence, we consider the range of $\delta$ to be $\{ 1, 100  \} $. Using the above parametrization in Eq.~(\ref{eq:baryon_Asym-M}), we get:
\begin{align}\label{eq:baryon_Asym-Parameter}
\frac{m^3 \times \delta^3 }{\gamma^3}  \approx  n \, 10^7 . 
\end{align}
\begin{figure}[!hbt]
	\centering
	\includegraphics[width=0.8\textwidth]{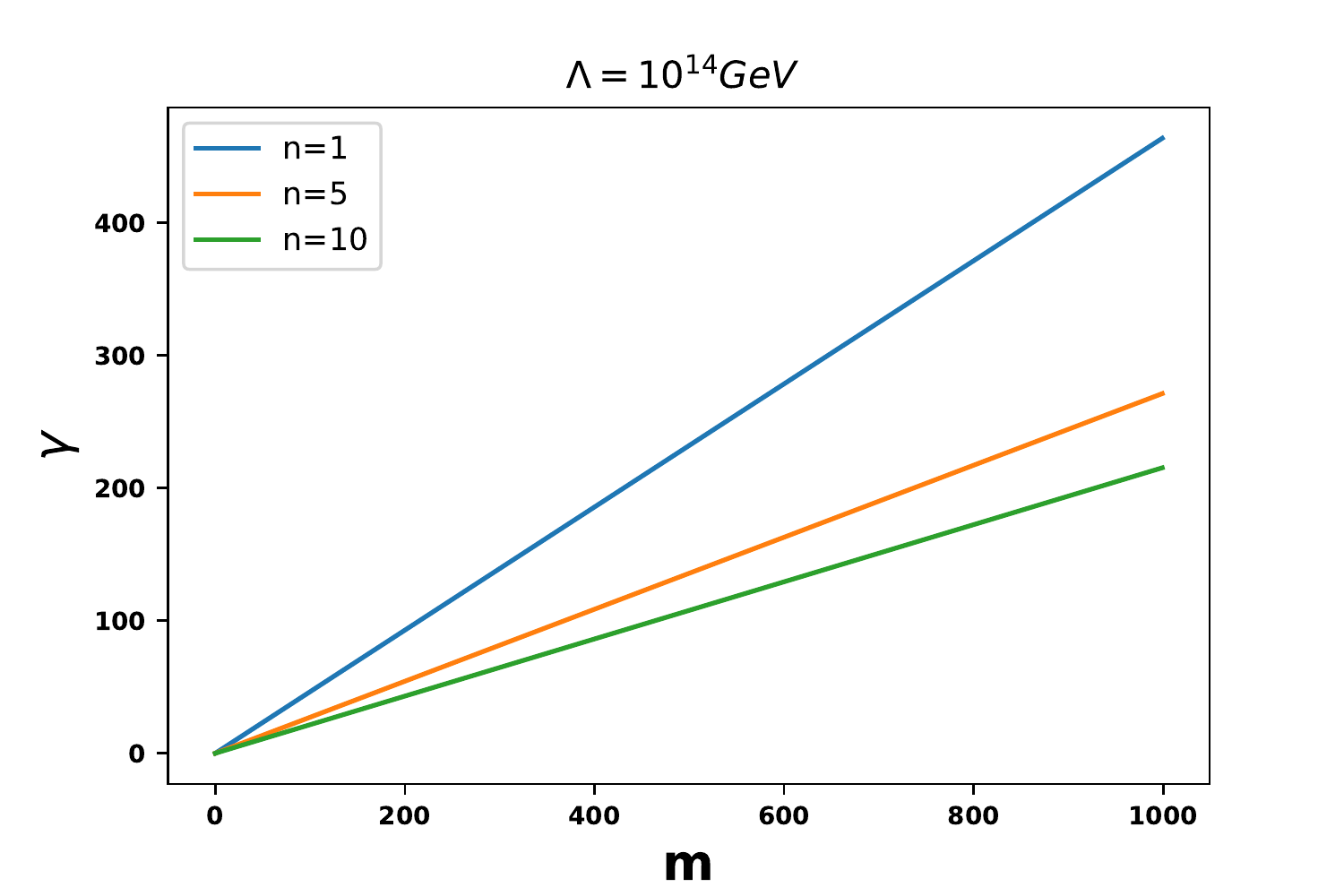}
	\caption{Plot of the rescaled reheating temperature $T_{RH}$  with the rescaled conformal symmetry breaking  parameter $M$, for different values of $n$. Here, we have set $\Lambda = 10^{14} GeV, \mu = 10^{10} GeV$.}
	\label{fig:Plot}
\end{figure}

\begin{figure}[!hbt]
\centering
\subfigure[]{%
\includegraphics[height=2in]{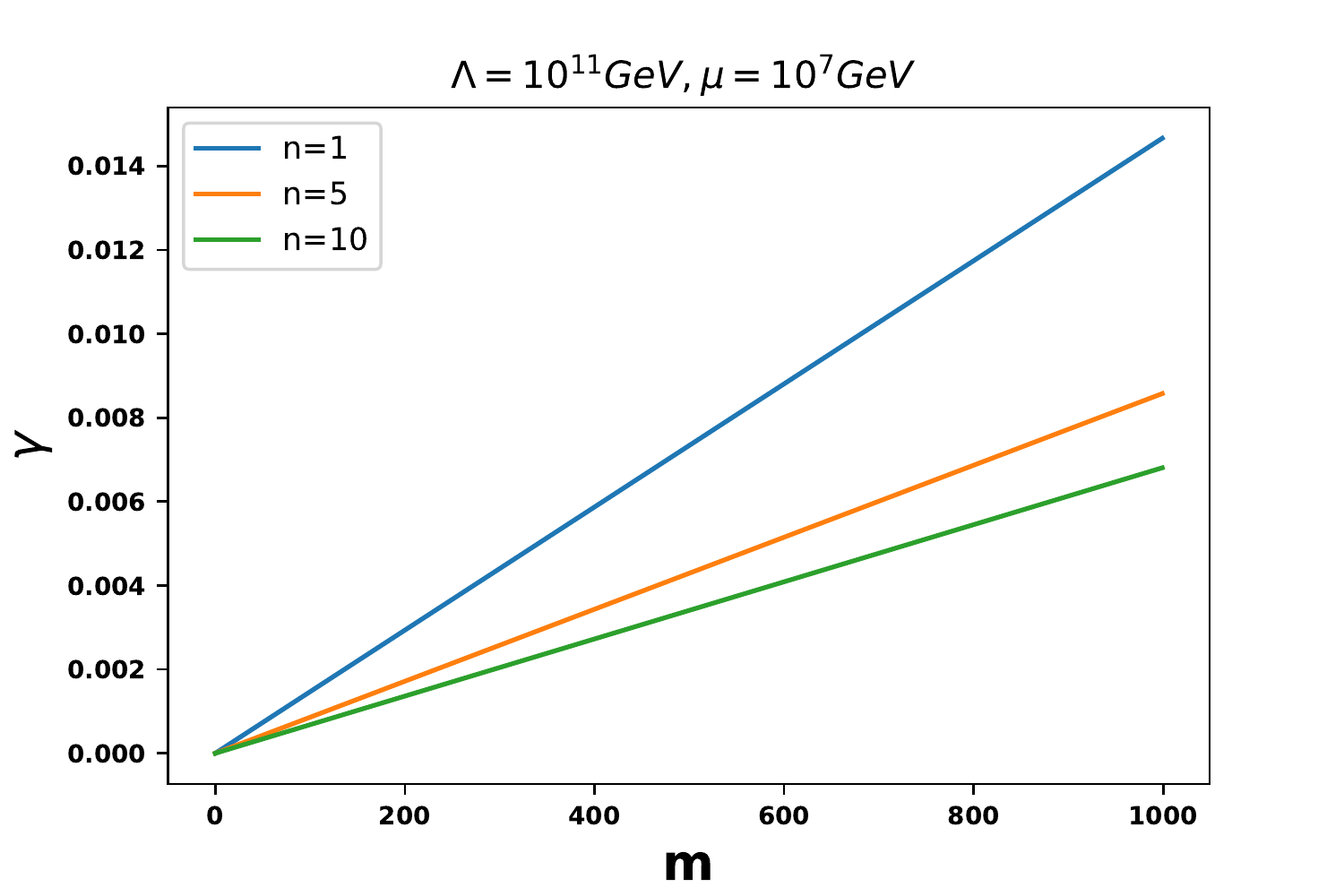}}%
\quad
\subfigure[]{%
\includegraphics[height=2in]{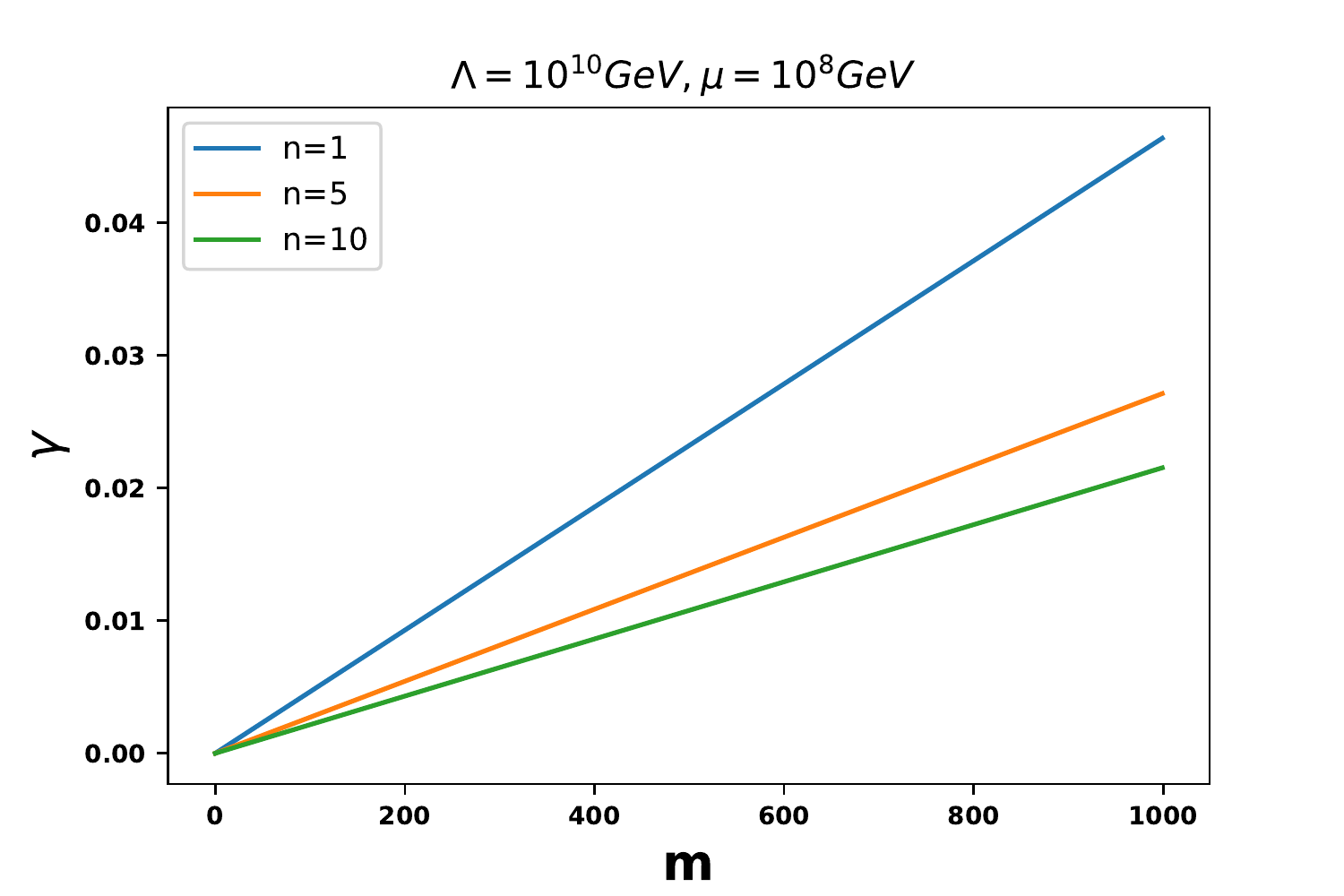}}%
\quad
\subfigure[]{%
\includegraphics[height=2in]{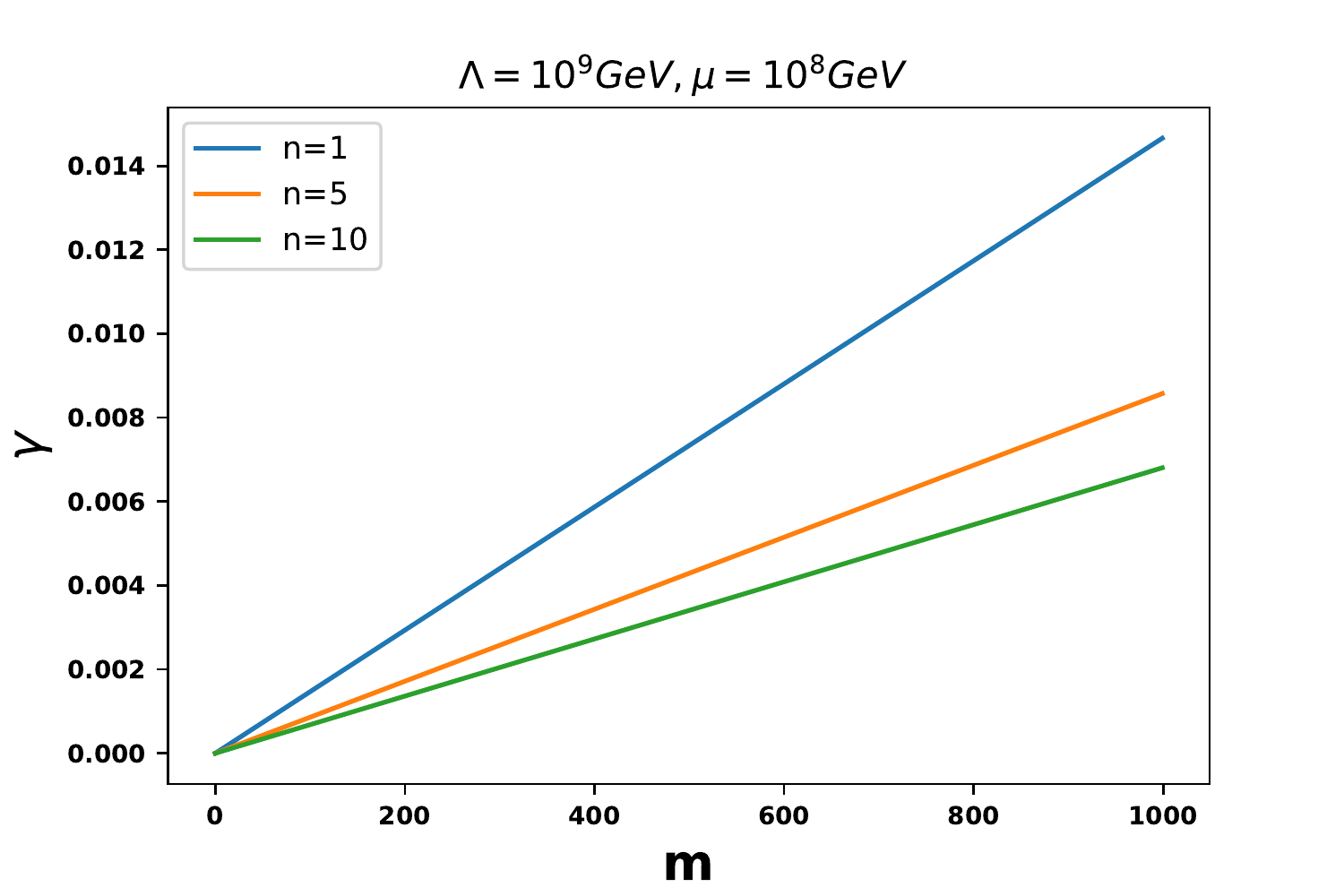}}%
\quad
\subfigure[]{%
\includegraphics[height=2in]{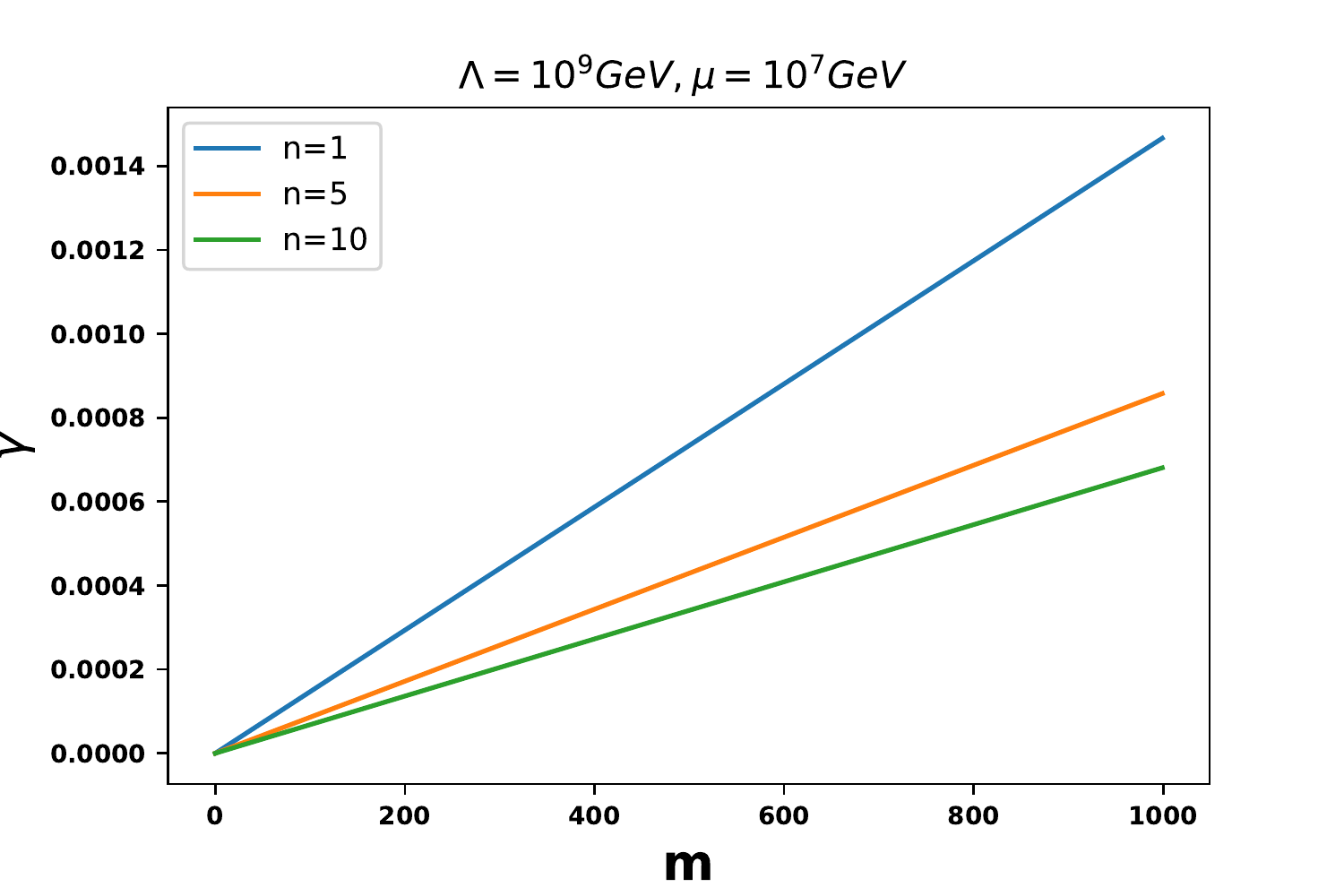}}%
\caption{Plots showing the behaviour of reheating temperature $T_{RH}$ (vertical axis) with parameter $M$ (horizontal axis), for lower energy scales 
of $\Lambda$ and $\mu$.}
\label{fig:reheatingGeneric2}
\end{figure}

\ref{fig:Plot} and \ref{fig:reheatingGeneric2} contain the plots of $\gamma$ versus $m$ for different values of $n$ and fixed $\delta$. In Appendix \eqref{app:Calculations} we have plotted the same for other values of $\delta$. From these plots, we deduce the following: First, for a range of values of 
$\gamma,$ $\delta$, and $m$, 
BAU can have values between $10^{-10}$ to $10^{-9}$. Thus, the model 
can lead to the observed amount of baryon asymmetry of the Universe consistent with the Planck data~\cite{2018-Planck}. Second, the model does not depend on the nature of the reheating dynamics. As can be seen from the plots, for a range of values of $m, \delta$, the model can lead to BAU for a range of reheating temperatures. This has to be contrasted with other models in the literature~\cite{2014-Long.Sabancilar.Vachaspati-JCAP,2016-Fujita.Kamada-PRD} which requires detailed knowledge of the reheating phase of the Universe. Third, the unknown parameter in the model is $M$. In Ref.~\cite{2020-Kushwaha.Shankaranarayanan-PRD}, we showed that for the model to be consistent with the lower limit of $10^{-16}$ Gauss magnetic fields in the voids~\cite{2010-Neronov.Vovk-Sci}, then $M \sim 10^{17} GeV$. The 
current analysis shows that $M \sim 10^{17} GeV$ is consistent with baryogenesis. Thus, the model is \emph{tantalizingly close} to solving baryogenesis and magnetogenesis using the same causal mechanism that solves the origin of density perturbations.

\section{Conclusions and Discussions}
\label{sec:conc}

In this work, we have proposed a viable baryogenesis scenario in the early Universe that does not require any extension to the Standard Model of particle physics. The crucial ingredient is the generation of primordial helical magnetic fields
due to Riemann coupling. The advantage of the primordial helical fields is that the non-zero helicity suggests a non-zero contribution in the CP violation term. 
An interesting feature of our model is the stretching of the primordial helical magnetic fields to super-horizon scales during inflation --- the same mechanism that leads to primordial density perturbations. While the helical modes generated around 40 - 60 e-foldings before the end of inflation lead to the observed large-scale magnetic fields, the helical modes that renter the horizon very early (at the beginning of the  radiation-dominated epoch) lead to the baryon asymmetry.  Thus, our mechanism provides \emph{possible testable evidence} for the entire inflationary epoch. 

More than two decades ago, Davidson pointed out an interesting relation between the primordial magnetic field and Sakharov's conditions~\cite{1996-Davidson-PLB}. In this work, we have explicitly shown that Davidson's conditions are necessary \emph{but not} sufficient. The key missing ingredient is the requirement of \emph{primordial helical magnetic fields}. While the helical and non-helical fields break the isotropy and lead to CP violation, only the modes with maximal helicity contribute significantly to the Chern-Simon number density. We have shown that the BAU parameter predicted by our model is independent of any specific inflation model and reheating dynamics; however, it depends on the scale at which inflation ends and reheating temperature. 
 
The BAU parameter \eqref{eq:baryon_Asym-M} obtained in our model is inversely proportional to reheating temperature. Assuming the exit of inflation at $10^{14}$~GeV, for the observed amount of baryon asymmetry $\eta_B \sim 10^{-10}$, we obtained  that the reheating temperature should be in the range 
$10^{12} - 10^{14}$~GeV, which is consistent with the constraints on the reheating temperature \cite{2006-Bassett.Tsujikawa.Wands-RevModPhys,1984-Ellis.Kim.Nanopoulos-PLB,1999-Benakli.Davidson-PRD}. 
This means that our model \emph{does not} prefer a very low-energy reheating temperature~\cite{1999-Benakli.Davidson-PRD}. 

In the literature, various mechanisms have been discussed to solve the BAU problem using the primordial helical magnetic fields~\cite{2014-Barrie.Kobakhidze-JHEP,2014-Long.Sabancilar.Vachaspati-JCAP,2016-Fujita.Kamada-PRD,2015-Anber.Sabancilar-PRD,2019-Domcke.etal-JCAP}. 
In Ref.~\cite{2016-Fujita.Kamada-PRD}, the authors obtained the required BAU by assuming the presence of helical magnetic fields of present-day strength $10^{-14} G < B_0 < 10^{-12} G $ and coherence length $1 \rm{pc} < \lambda < 1 \rm{Mpc}$, and taking into account of the MHD effects. In Ref.~\cite{2014-Long.Sabancilar.Vachaspati-JCAP}, authors studied the generation of a primordial magnetic field in conjunction with the BAU generation through leptogenesis; however, the predicted value of the present-day coherence length of such magnetic fields is very small$\sim 10$~pc.

In Refs.~\cite{2015-Anber.Sabancilar-PRD,2019-Domcke.etal-JCAP}, the authors consider pseudoscalar inflation (axion inflation) model with a dimension five couplings. In these models, the authors assumed the scale of the baryogenesis to be electroweak scale, and they obtained the required BAU assuming the scale of inflation to be $10^{10} \rm{GeV}$ --- $10^{12} \rm{GeV}$~\cite{2019-Domcke.etal-JCAP,2015-Anber.Sabancilar-PRD}.  In Ref. ~\cite{2014-Barrie.Kobakhidze-JHEP}, the authors considered the extension of the Standard Model with anomalous gauge symmetry. They obtained the 
required BAU for $H_{Inf} \sim 10^{14} \rm{GeV}$ and reheating temperate at $10^{16}\rm{GeV}$. In Ref.~\cite{1999-Brustein.Oaknin-PRL}, the authors argued that to generate the observed baryon asymmetry, some asymmetry in the initial conditions of either $\bf{B}$ or scalar field $\phi$ is required, which can be induced from temperature-dependent potential or asymmetry in quantum fluctuations. Our model is robust to inflationary/reheating dynamics and uses the same success of inflationary perturbations to generate BAU. Thus, our model is \emph{tantalizingly close} to solving baryogenesis and magnetogenesis using the same causal mechanism that solves the origin of density perturbations.

In this work, we did not consider the gravity contribution to the chiral anomaly equation. In Ref.~\cite{2006-Alexander.Peskin.Jabbari-PRL}, the authors considered the phenomenon of gravitational birefringence to show that the gravitational fluctuations generated during inflation can give the Universe's observed amount of baryon asymmetry. However, as we showed in. 
Sec.~\eqref{sec:Baryo-magnetic}, $R\tilde{R}$ contributes only in the second-order, and hence we have ignored it in this analysis. It may be interesting to look at the second-order corrections and analyze the parameter constraints.

In this work, we have used the general effective field theory of gravity coupled to the Standard Model of particle physics framework to obtain leading order gravity terms that couple to the standard model Bosons~\cite{2019-Ruhdorfer.etal-JHEP}.  We have considered only the mass dimension 6-operators coupling to the gauge field Lagrangian, specifically, to the electromagnetic field. 
The coupling to the Fermions arises at the mass dimension 8. Thus, coupling of Fermion-anti-Fermion with $U(1)$ field will play a role only at this order. While these are expected to be suppressed compared to mass-dimension 6 operators, they are relevant at Planck scale. We plan to look at the effects of mass dimension 8 operators on the baryogenesis.

In this work, we focused on the electromagnetic fields and the effects of the helical fields on baryogenesis. It will be interesting to extend the analysis to 
Gluons and study the effects on the asymmetry generated in quarks and the Baryons. It is particularly important, and a study on this is currently
in progress to acquire more stringent constraints on the parameters $M$ and $T_{\rm RH}$~\cite{2021-Sharma.Ashu.Shanki}. 

\noindent  {\it Note added:}  As we were finalizing this manuscript, the article \cite{2021-Giovannini-arXiv} appeared on the arXiv which also discusses Baryogenesis from Magnetic fields. However, the approach followed in the reference requires MHD amplification while our approach requires helical fields generated during inflation.

\begin{acknowledgments}
The authors thank Joseph P. Johnson and Urjit A Yajnik for comments on the earlier version of the manuscript. The authors thank Kandaswamy Subramanian for useful discussion. The authors thank the anonymous referee for raising some points which clarified important issues in the work. The MHRD fellowship at IIT Bombay financially supports AK. This work is supported by the ISRO-Respond grant. 
\end{acknowledgments}

\appendix

\section{Quantization in the Helicity basis}
\label{app:helicity_basis}
In this section, we briefly discuss the evolution of the quantum fluctuations of the electromagnetic field in the helicity basis~~\cite{2018-Sharma.Subramanian.Seshadri.PRD}.  Decomposition of the vector potential in Fourier domain leads to:
\begin{align}\label{eq:FourierT}
A^{i}(\vec{x}, \eta) =  \int \frac{d^3 k}{(2\pi)^3} \sum_{\lambda = 1,2} \varepsilon^i_{\lambda} \left[ A_{\lambda}(k,\eta) b_{\lambda}(\vec{k}) e^{ik\cdot x}  
+ A^*_{\lambda}(k,\eta)  b^{\dagger}_{\lambda}(\vec{k}) e^{- ik\cdot x} \right]
\end{align}
where $b(\textbf{k})$ and $b^{\dagger}(\textbf{k})$ are the annihilation and creation operators respectively for a given comoving mode $\textbf{k}$, and $\varepsilon_{\lambda}^i$ is the orthogonal basis vector which in right-handed coordinate system~\cite{2018-Sharma.Subramanian.Seshadri.PRD} is given by
\begin{align}\label{eq:basisVector}
\varepsilon^{\mu} = \left( \frac{1}{a}, \textbf{0} \right), \,\,\,\, \varepsilon^{\mu} = \left( 0, \frac{ \hat{\varepsilon}^i_{\lambda} }{a} \right), \,\,\,\, \varepsilon^{\mu}_3 = \left(  0, \frac{\hat{\textbf{k}}}{a} \right) \quad  \text{for} \quad \lambda = 1, 2 \, ,
\end{align}
3-vectors $\hat{\varepsilon}^i_{\lambda}$ are unit vectors orthogonal to $\hat{\textbf{k}}$ and to each other. Substituting Eq.~(\ref{eq:basisVector}) in 
Eq.~(\ref{eq:FourierT} ) and defining the new variable 
$\bar{A}_{\lambda} = a(\eta) \,  A_{\lambda}(k,\eta)$, we have:
\begin{align}\label{eq:Fdecomposition}
A_{i}(\textbf{x}, \eta) = \int \frac{d^3 k}{(2\pi)^3} \sum_{\lambda = 1,2} \,\hat{\varepsilon}_{i \lambda} \left[ \bar{A}_{\lambda} b_{\lambda}(\textbf{k}) e^{ i \textbf{k} \cdot \textbf{x} }  
+ \bar{A}^*_{\lambda}  b^{\dagger}_{\lambda}(\vec{k}) e^{- i \textbf{k} \cdot \textbf{x} } \right] \, .
\end{align}
Substituting  Eq.~\eqref{eq:Fdecomposition} in Eq.~\eqref{eq:equation_of_motion}, we get:
\begin{align}\label{eq:EOM_fourier_space}
\sum_{\lambda = 1,2}b_{\lambda} \left[  \hat{\varepsilon}_{i \lambda}  \bar{A}_{\lambda}^{\prime\prime} + \frac{4i}{M^2} \epsilon_{i j l} k_j \hat{\varepsilon}_{l \, \lambda} \bar{A}_{\lambda} \, \left( \frac{a^{\prime\prime\prime} }{a^3} - 3\frac{ a^{\prime\prime}  a^{\prime}  }{a^4} \right) +  k^2 \hat{\varepsilon}_{i \lambda} \bar{A}_{\lambda}\right] = 0 
\end{align}
where we have used $\partial_j \partial_j = -k^2$. 

Since the action \eqref{eq:action} contains parity breaking term (helicity term), it is useful to work in the helicity basis. The helicity basis vectors $\varepsilon_+$ and $\varepsilon_-$ corresponding to $h = +1$ and $h = -1$ are defined as
\begin{align}\label{eq5:helicity_basis}
\varepsilon_{\pm} = \frac{1}{\sqrt{2}} \left(   \hat{\varepsilon}_1 \pm i  \hat{\varepsilon}_2  \right).
\end{align}
Assuming that the wave propagates in the $z-$direction, the vector potential in the helicity basis is given by:
%
\begin{align}\label{eq5:decomp_A_helicity}
\bar{\textbf{A}} = \bar{A}_1 \hat{\varepsilon}_1 + \bar{A}_2  \hat{\varepsilon}_2 = A_+ \varepsilon_+ + A_- \varepsilon_-
\end{align}
where $A_+$($A_-$) refer to the vector potential with positive (negative) helicity. 
The ground state in the helicity basis is defined as
\begin{align}\label{eq:GS_helicity}
b_h(\textbf{k}) | 0 \rangle = 0 
\end{align}
and satisfy the following commutation relations:
\begin{align}\label{eq:comm-b_h}
\left[ b_h(\textbf{k}), b^{\dagger}_{h^{\prime}}(\textbf{q})  \right] &= \left( 2\pi \right)^3 \, \delta^3(\textbf{k} - \textbf{q}) \, \delta_{h h^{\prime}} \\
\left[ b_h(\textbf{k}), b_{h^{\prime}}(\textbf{q})  \right] &= 0 = \left[ b^{\dagger}_h(\textbf{k}), b^{\dagger}_{h^{\prime}}(\textbf{q})  \right] \, .
\end{align}

Rewriting \eqref{eq:EOM_fourier_space} in the Helicity basis and replacing
$\epsilon_{i j l} \partial_j A_l \longrightarrow  -k \sum_{h = \pm 1} h A_h \varepsilon_{h}$, we have:
\begin{align}\label{App_eq:eom_helicity}
A_h^{\prime\prime} + \left[  k^2 - \frac{4kh}{M^2} \, 
\Gamma(\eta)  \right] A_h= 0 \, ,
\end{align}
where,
\begin{equation}
\label{def:Gamma}
  \Gamma(\eta) = \frac{a^{\prime\prime\prime}}{a^3} - 3\frac{a^{\prime\prime} a^{\prime} }{a^4} = \frac{1}{a^2} \left(\mathcal{H}'' - 2 \mathcal{H}^3\right) \, .  
\end{equation}
%
%
%
%
%

\section{Generation and evolution of helical modes }
\label{app:Helical}

Substituting the power-law inflation scale factor (\ref{eq:powerLaw}) in 
Eq.~(\ref{eq:eom_helicity}), we have:
\begin{align}\label{eq:arbitrayBeta_eta}
{A_h^{\prime\prime} + \left[ k^2 - \frac{8kh}{M^2} 
\frac{ \beta (\beta+1) ( \beta + 2)}{\eta_0^3} 
\left( \frac{-\eta_0}{\eta} \right)^{(2 \beta+5)} \right]  \, A_h = 0 } \, .
\end{align}
Helicity term  vanishes for de-sitter case ($\beta = -2$), which is consistent with the fact that the de Sitter symmetry will not be preserved in the presence of helicity terms. However, it will be non-zero for the approximately de Sitter universe i.e., $\beta = -2-\epsilon$. 
In sub horizon limit ($\left| - k \eta \right| \gg 1$), Eq.~(\ref{eq:arbitrayBeta_eta}) simplifies to:
\begin{align}\label{eq:sub-horizon}
A_h^{\prime\prime} + k^2 A_h \approx 0 
\end{align}
and assuming that the quantum field is in the vacuum state at asymptotic past 
(Bunch-Davies vacuum state), we have:
\begin{align}
A_h = \frac{1}{\sqrt{k}} e^{-ik\eta}.
\end{align}
On super-Horizon scales ($\left| - k \eta \right| \ll 1$), Eq.~(\ref{eq:arbitrayBeta_eta}) becomes:
\begin{align}\label{eq:sup_mode_alpha_tau-varsigma}
{   \alpha^2 \frac{d^2 A_h}{d\tau^2} + \frac{\alpha(\alpha+1)}{\tau} \frac{d A_h}{d \tau} + h \, k \, \varsigma^2 A_h  = 0     }
\end{align}
where  
\begin{equation}
\label{def:varsigma}
{\varsigma^2 \equiv -\frac{1}{M^2 \, \eta_0} (2\alpha - 3)(2\alpha - 1)(2\alpha + 1)\, ,
\tau = \left( -\frac{\eta_0}{ \eta} \right)^{\alpha}  \, ,
\alpha = \beta + \frac{3}{2} }
\end{equation}
Note that $\tau$ (dimensionless variable $0 < \tau < \infty$) and $\eta$ (negative during inflation) are linearly related. [At the start of inflation, $\tau$ is large and vanishes 
at the end of inflation.] Note that $\alpha = -\frac{1}{2}$ corresponds to de-sitter and $ \alpha \leq - \frac{1}{2} $.
The solutions for the above equation (\ref{eq:sup_mode_alpha_tau-varsigma}) are:
\begin{subequations}
\begin{align}\label{eq:sup_mode_h+}
A_{+}(\tau,k) &= \tau^{- \frac{1}{2\alpha} } \, J_{  \frac{1}{2 \alpha}} \left( \frac{\varsigma \sqrt{k} }{\alpha} \tau \, \right)  C_1+ \tau^{- \frac{1}{2\alpha} } \, Y_{ \frac{1}{2 \alpha} }  \left( \frac{ \varsigma \sqrt{k} }{\alpha}\tau \right)  C_2
\\
\label{eq:sup_mode_h-}
A_{-}(\tau,k) &=  \tau^{- \frac{1}{2\alpha} } \, J_{  \frac{1}{2 \alpha}} \left(  -i \frac{ \varsigma \, \sqrt{k} }{\alpha} \tau  \right)  C_3+ \tau^{- \frac{1}{2\alpha} }  \, Y_{  \frac{1}{2 \alpha} } \left(  - i \, \frac{ \varsigma\, \sqrt{k} }{\alpha} \tau  \right)  C_4 \, ,
\end{align}
\end{subequations}
where $C_1, C_2, C_3, C_4$ are arbitrary constants of dimension $L^{1/2}$. 
For the two helicity modes, we fix the constants $C_1, C_2$ ($C_3, C_4)$ by matching $A_h$ and $A_h'$ at the transition time of sub-horizon and super-horizon modes at $k_* \sim \eta_*^{-1}$ where $*$ refers to the 
quantities evaluated at the horizon-exit.

Although the analysis can be done for any general value of $\alpha$, to keep the calculations tractable, we obtain the constants for $\alpha = -1$.  There are two reasons for this choice: First, in this special case, $\tau \propto \eta$ and the super-horizon modes can be written in terms of $\eta$ using the linear relation. 
Second, the constants $C_1, C_2, C_3, C_4$ have a weak dependence of $\alpha$ 
and, hence, finding the value for a given value of $\alpha$ will be accurate within an order~\cite{2020-Kushwaha.Shankaranarayanan-PRD}. Thus, matching the solutions and the derivatives at the horizon-exit, we get:
\begin{align}\label{eq:Coefficients}
C_1 &= -e^i \,  \sqrt{ \frac{\pi \eta_0}{ 2} } \left( \frac{1}{\sqrt{\Theta}} \rm{sin}\Theta 
 + i \sqrt{ \Theta }  \,  \rm{cos} \Theta   \right), \,\,\,\,
C_2 = -i \, e^i \,  \sqrt{ \frac{\pi \eta_0}{ 2} } \left( \frac{1}{\sqrt{\Theta}} \rm{cos}\Theta 
 - i \sqrt{ \Theta }  \,  \rm{sin} \Theta   \right) \\
C_3 &= e^i \,  \sqrt{ \frac{\pi \eta_0}{ 2 } } \left( \frac{1}{\sqrt{i \Theta}} \rm{sinh}\Theta 
 +  \sqrt{ i \Theta }  \,  \rm{cosh} \Theta   \right), \,\,\,\,
C_4 = -i \, e^i \,  \sqrt{ \frac{\pi \eta_0}{ 2 } } \left( \frac{1}{\sqrt{i \Theta}} \rm{cosh} \Theta
 +  \sqrt{ i \Theta }  \,  \rm{sinh} \Theta   \right). \nonumber
\end{align}
where $\Theta = \sqrt{ \frac{15 \eta_*}{M^2 \eta_0^3} }$ is the dimensionless constant.

In Ref.~\cite{2020-Kushwaha.Shankaranarayanan-PRD}, 
the current authors derived the magnetic field spectral energy density and is given by
\begin{align}\label{eq:powerSpetrum}
\frac{d\rho_B}{d\rm{ln}k}   =  \left| {\cal C}(k_*, \alpha) \right|^2 \, \left( \frac{ k }{k_*} \right)^{ 2 - 4\alpha }   k^{3 + 4\alpha +\frac{1}{2\alpha}}  
 +  \left| {\cal C}_2(k_*,\alpha)  \right|^2 \,
   \left( \frac{ k }{k_*} \right)^{ 4 - 4\alpha }   k^{1 + 4\alpha - \frac{1}{2\alpha}}  
\end{align}  
\begin{figure}[!hbt]
\centering
\subfigure[]{%
\label{fig:Powerfirst}%
\includegraphics[height=2in]{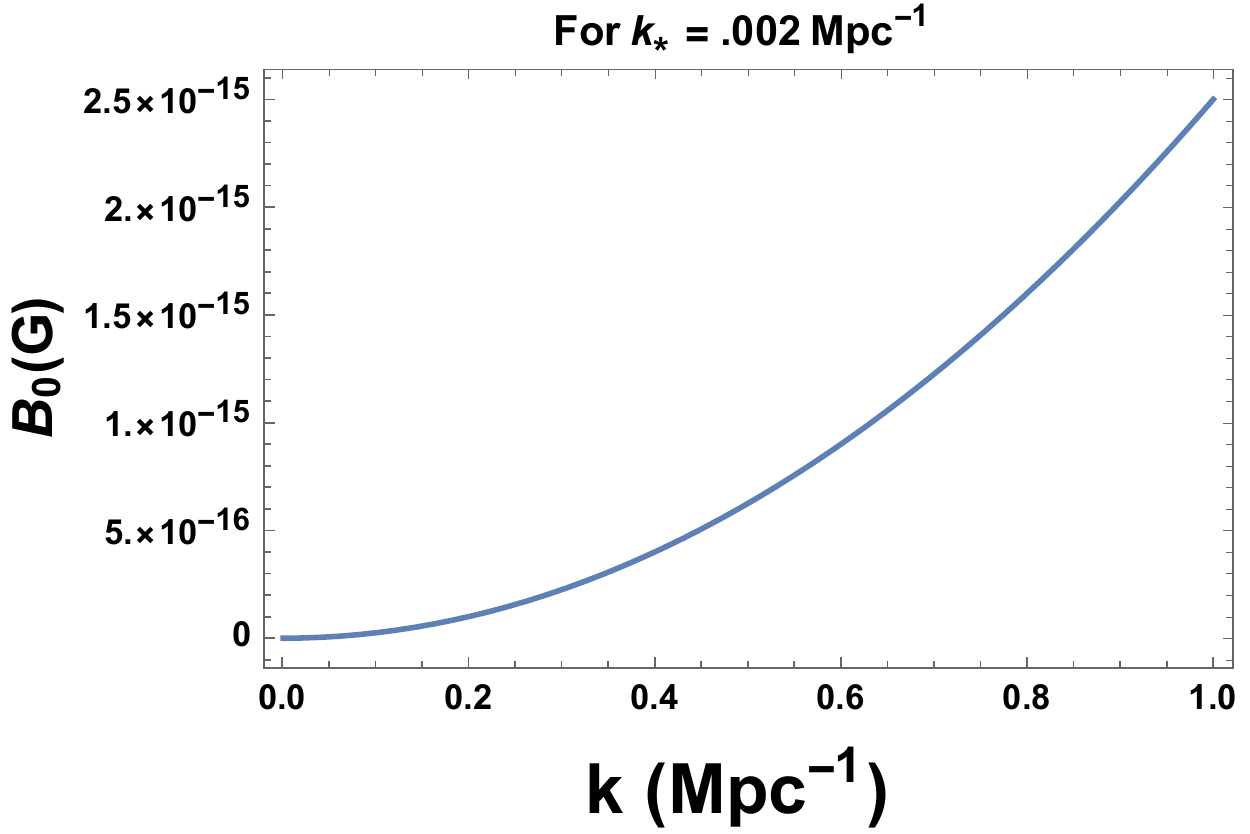}}%
\quad
\subfigure[]{%
\label{fig:Powertsecond}%
\includegraphics[height=2in]{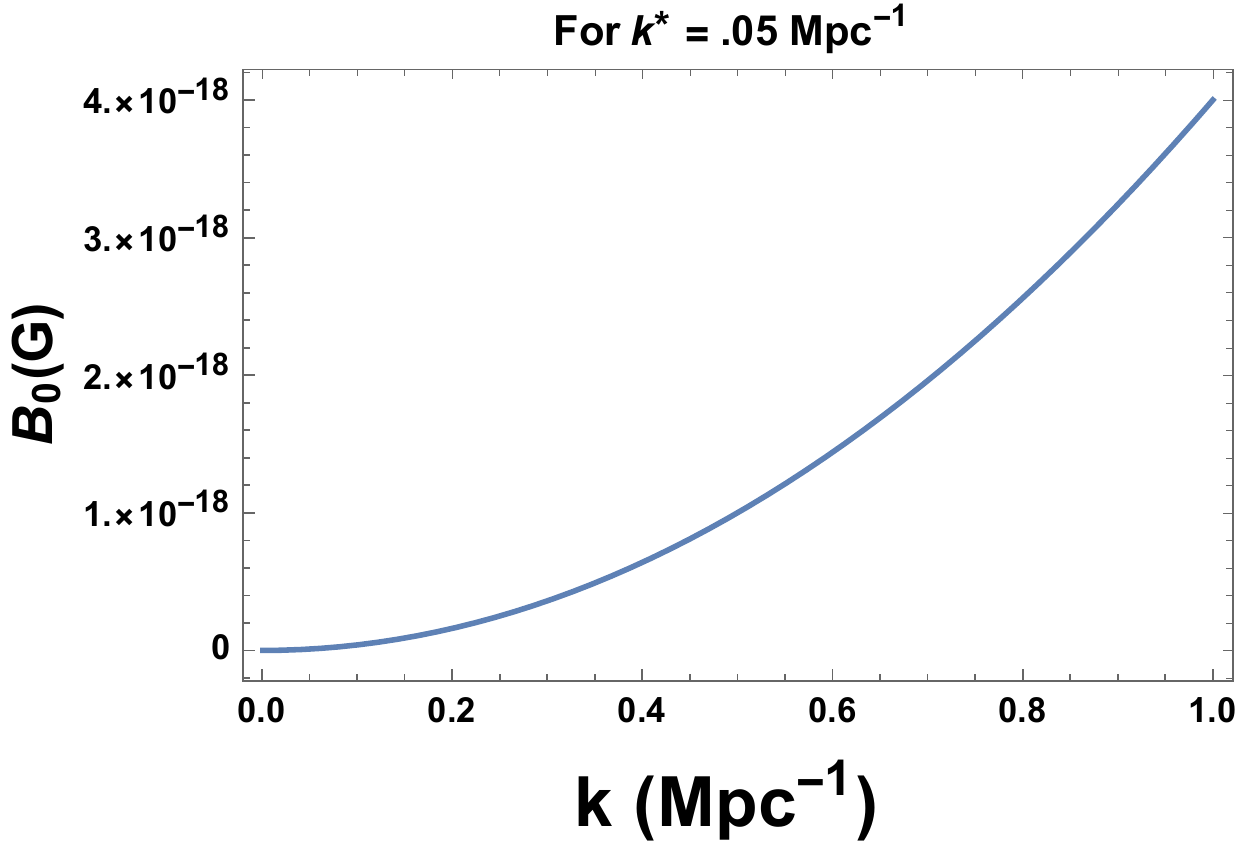}}%
\caption{ Plot showing the strength of the present day primordial helical magnetic field at different length scales (from Mpc to Gpc), for two different pivot scales.}
\label{fig:PowerSpectrum}
\end{figure}
where $k_*$ is the pivot scale, and 
${\cal C}(k_*, \alpha)$ and ${\cal C}_2(k_*, \alpha)$ are constants that depend on the inflationary energy scale (See Eq. 44 in Ref. \cite{2020-Kushwaha.Shankaranarayanan-PRD}). For de Sitter inflation ($\alpha = - 1/2$), 
the present day magnetic field as a function of $k$ is given by:
\begin{equation}
B_0(k) \sim 10^{-20} \left( \frac{k}{k_*} \right)^2 \, \rm{G} 
\end{equation}
where we have included only the leading order contribution and have discarded the subleading contribution.

In \ref{fig:PowerSpectrum} we have plotted the power spectrum of the present day primordial helical magnetic field at different length scales for two pivot scales $k_* = 0.002 \, Mpc^{-1}$~\cite{Liddle.Lyth-Book} and $k_* = 0.05 \, Mpc^{-1}$~\cite{2018-Planck}. One can see from figure \ref{fig:Powerfirst} that for around Mpc scale, the value of present day magnetic field is $10^{-15} G$. This is consistent with the current observations~\cite{2010-Kahniashvili.Ratra.etal-PRD,2013-Durrer.Neronov-Arxiv,2016-Subramanian-Arxiv,2004-Giovannini-IJMPD,2020-Vachaspati-arXiv}.

Using the fact that modes exit the horizon around 5 e-foldings, 
\begin{align}\label{eq:eta_*}
\eta_* = \eta_{end} \cdot 10^2
\end{align} 
and $\mathcal{H} \sim {\eta_0}^{-1} \sim 10^{14} \rm{GeV} $, for $M \sim 10^{14} -10^{17} \rm{GeV}$~\cite{2020-Kushwaha.Shankaranarayanan-PRD,2004-Shankaranarayanan.Sriramkumar-PRD}, we obtain 
\[
\Theta  \approx  \sqrt{ \frac{\eta_{end} \cdot 10^{45} GeV^3  }{M^2 }} \, , 
\]
which is very small value.
Note also that 
\[
\eta_{end} = -\frac{1}{a(\eta_{end}) H} = -\frac{e^{-N_{\rm Inf}}}{H_{\rm Inf}} \approx 10^{-41} \, {\rm GeV}^{-1} \, .
\]
Using the fact that $\Theta$ is very small, we get
 \begin{align}\label{eq:ApproxCoefficients}
|C_1| \approx | C_3| \approx \sqrt{\Theta \, \eta_0} \,\, , \qquad \text{and} \qquad 
|C_2| \approx | C_4| \approx \sqrt{ \frac{\eta_0}{\Theta}}.
\end{align}
Hence, we obtain the following relations among the coefficients $|C_1| \approx |C_3| << |C_2| \approx |C_4|$.

\section{BAU parameter for arbitrary values of $\Lambda$ and $\mu$}
\label{app:Calculations}

\begin{figure}[!hbt]
\centering
\subfigure[]{%
\includegraphics[height=2in]{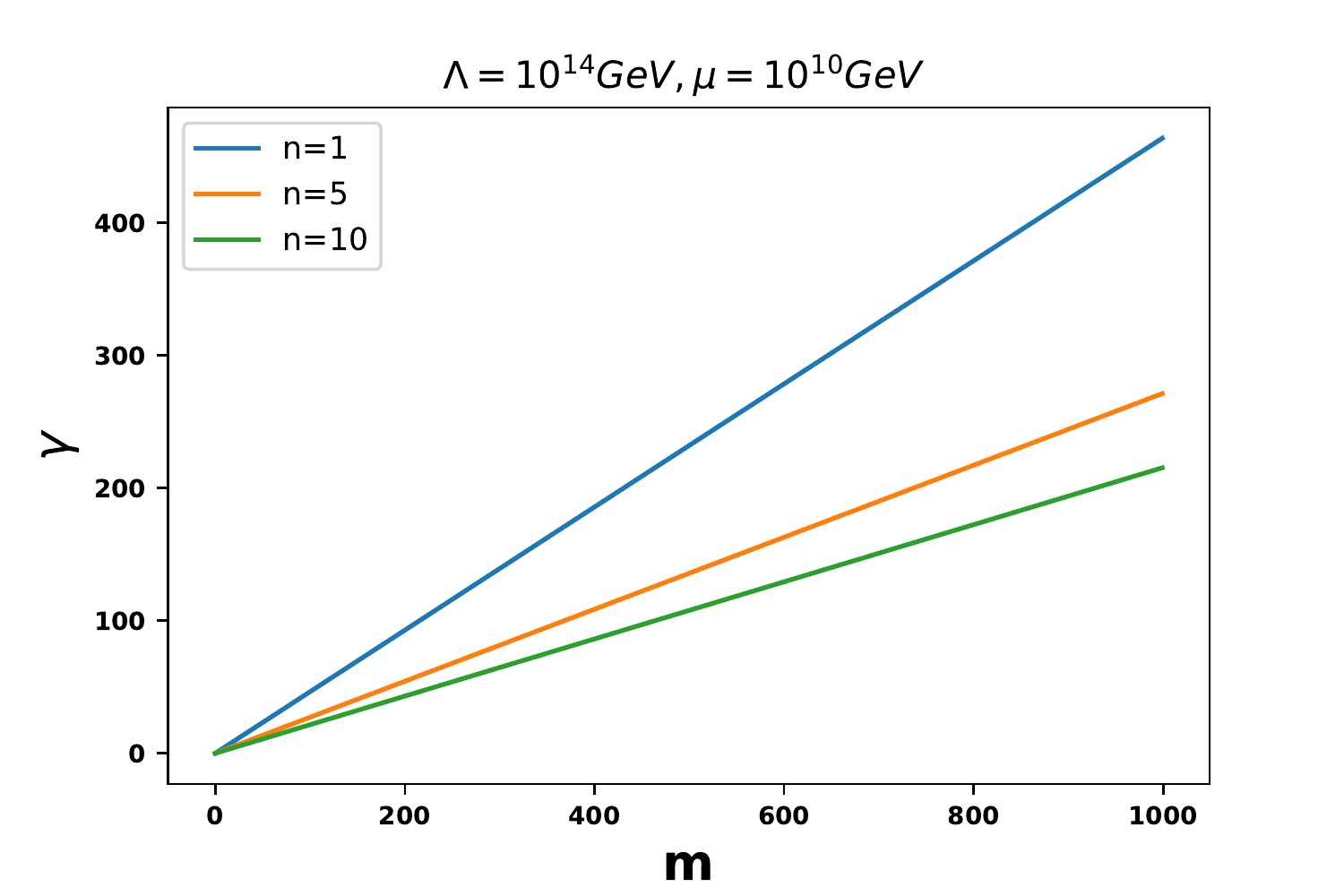}}%
\quad
\subfigure[]{%
\includegraphics[height=2in]{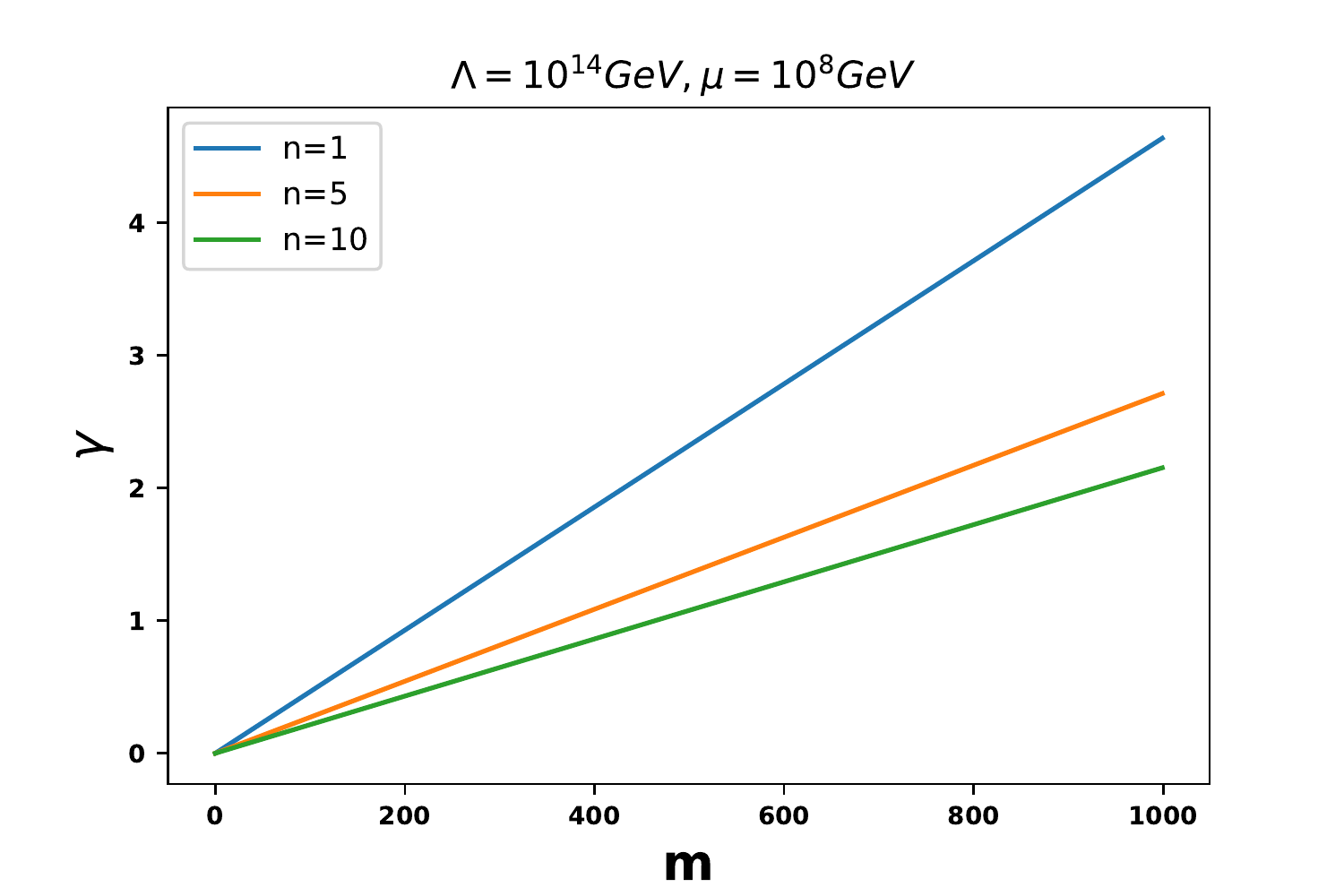}}%
\quad
\subfigure[]{%
\includegraphics[height=2in]{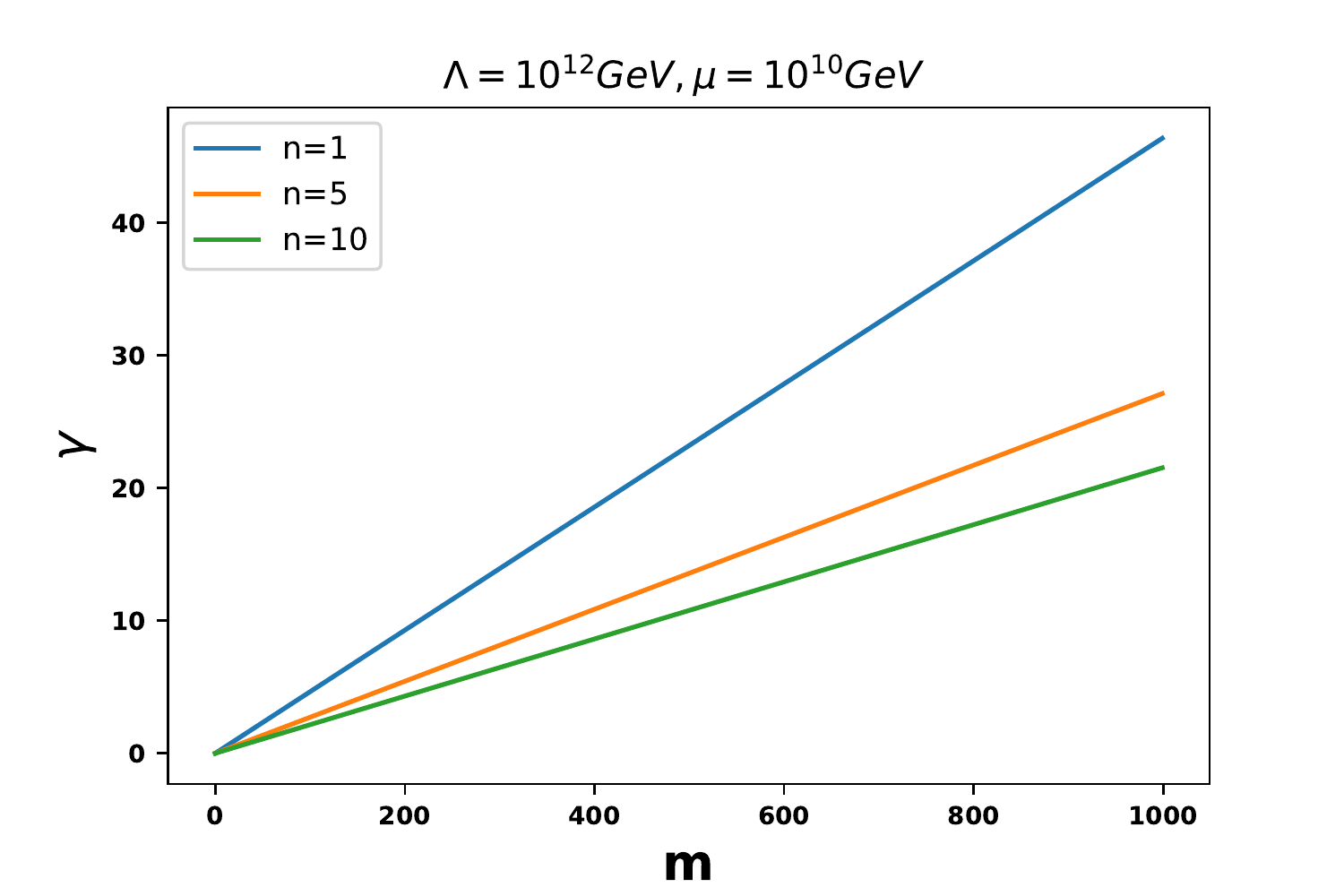}}%
\quad
\subfigure[]{%
\includegraphics[height=2in]{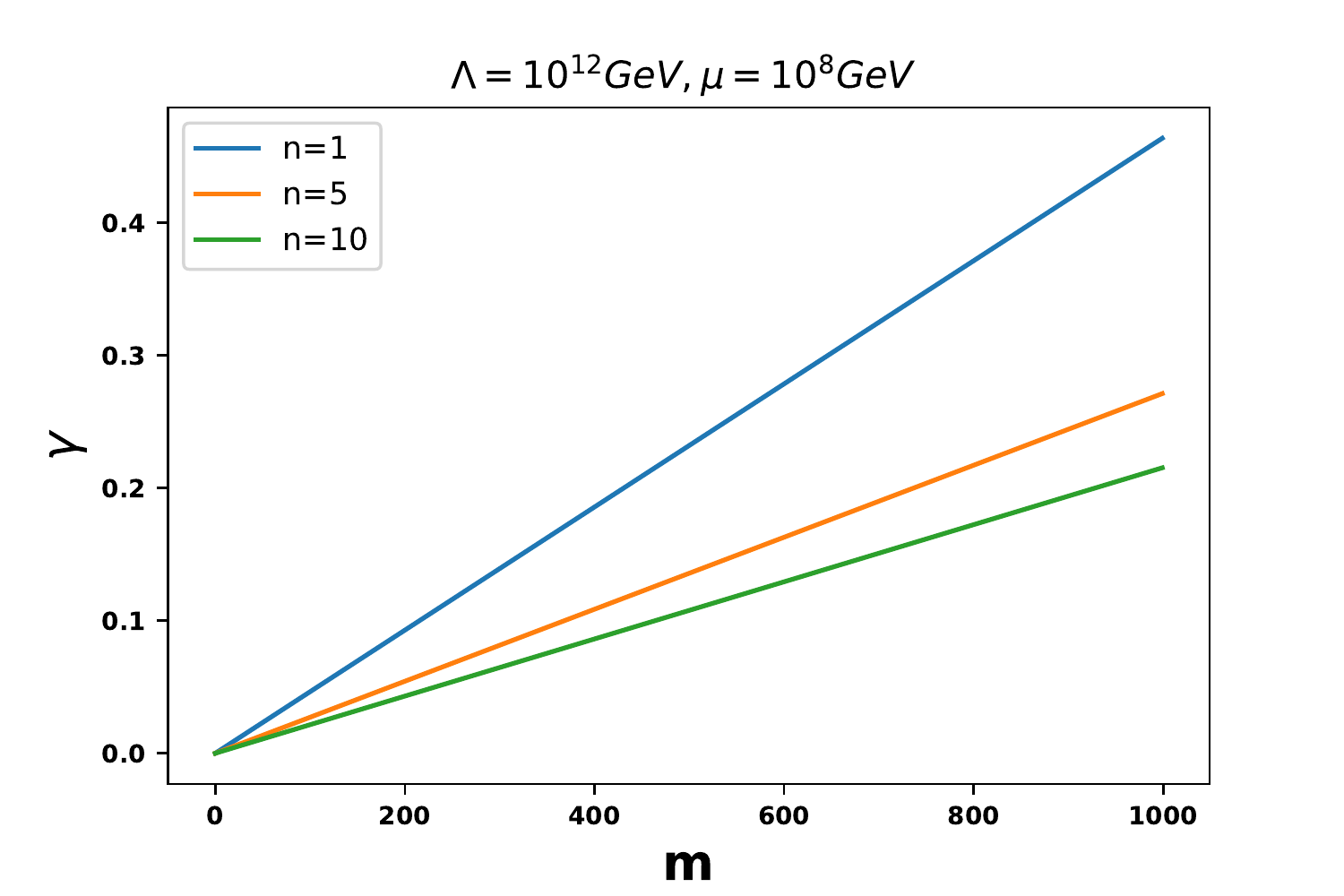}}%
\caption{ Plot showing the behaviour of reheating temperature $T_{RH}$ (vertical axis) with parameter $M$ (horizontal axis), for different ranges of $\Lambda,\mu$. In upper panel ( and lower panel) $\Lambda$ is fixed at $10^{14} GeV$ ( and $10^{12} GeV$) and lower energy scale is varied.}
\label{fig:reheatingGeneric}
\end{figure}
Following Ref.\cite{2016-Subramanian-Arxiv}, we first evaluate the contribution of dilution factor for arbitrary values of $\Lambda$ and $\mu$. Assuming instantaneous reheating, the universe transited to radiation domination after inflation.  Using entropy conservation i.e. $g a^3T^3 = \mbox{constant}$ during its evolution, where $g$ is effective relativistic degrees of freedom, we have: 
\begin{equation}
\frac{a_{\mu}}{a_{\Lambda}} =\left( \frac{g_{\Lambda}}{g_{\mu}}\right)^{1/3} \frac{T_{\Lambda}}{T_\mu}
\label{eq:AppC01}
\end{equation}
where $a_{\Lambda}, a_{\mu}$ are the scale factors at which the helical modes with energy $\Lambda$ and $\mu$ reentered the radiation dominated Universe, $g_{\Lambda}, g_{\mu}$ are the effective relativistic degrees of freedom at which the helical modes with energy $\Lambda$ and $\mu$ reentered the radiation dominated Universe, and $T_{\Lambda}, T_{\mu}$ are the Universe temperatures at which the helical modes with energy $\Lambda$ and $\mu$ reentered the radiation dominated Universe.  The Friedmann equation is: 
\begin{align}
\label{eq:AppC02}
H^2 = \frac{1}{3 M_{\rm P}^2} \left[ g_{\Lambda} \left(\frac{\pi^2}{30} \right) T_{\Lambda}^4 \right] \, .
\end{align}
The baryogenesis occurs as soon as the helical modes reenter the Hubble radius during the radiation-dominated epoch. For simplicity, we assume that baryogenesis occurs at the start of radiation-dominated epoch, hence we take energy scale of $\Lambda$ to be of order $H$. Substituting $T_{\Lambda}$ from Eq. \eqref{eq:AppC02} in Eq. \eqref{eq:AppC01}, we get,
  \begin{align}
 \frac{a_{\mu}}{a_{\Lambda}} = \frac{g_\Lambda^{1/12}}{g_\mu^{1/3}} \left( \frac{90}{\pi^2} \right)^{1/4} \frac{M_{\rm P}^{1/2} \Lambda^{1/2}}{T_\mu}  \approx \frac{ 10^{9} \sqrt{\Lambda}}{\mu},
 \end{align}
where $g_\Lambda \sim 100$ (during reheating) and $g_\mu$ is of the order 10. Physically, we see that the Universe has expanded by a factor of $\frac{10^9 \, \sqrt{\Lambda}}{\mu}$ when the helical modes of energy $\Lambda$ and $\mu$ reenter the radiation dominated epoch and hence, 
in Eq.~(\ref{eq:n_B}), the inverse of this factor will act as the dilution factor. 

Setting $\Lambda = 10^{14}$~GeV, and 
$\mu = 10^{10}$~ GeV, in the above expression, we have: 
\[
\frac{a_{\mu}}{a_{\Lambda}} = \frac{g_f^{1/12}}{g_1^{1/3}} \left( \frac{90}{\pi^2} \right)^{1/4} \frac{\left( M_{\rm P} \Lambda\right)^{1/2}}{\mu}  \approx 10^{6} \, .
\]
For the modes that exit the horizon during inflation at $N=5$, we have: 
\[ 
\frac{1}{a^3} \sim 10^{-33} \frac{\mu^3}{\Lambda^{3/2}} .
\]
Therefore, for generic scales of baryogenesis, the BAU parameter \eqref{eq:baryon_Asym} is given by
\begin{align}\label{eq:baryon_Asym-M-mu}
\eta_B  \approx  \frac{ 10^{-38} \cdot \eta_0^2 }{ \sqrt{\eta_{end} \cdot 10^{45} GeV^3} } \frac{M^3  \left( \Lambda^3 \mu^3 - \mu^6 \right)  }{ \Lambda^{3/2} \,  T^3_{\rm{RH}}  } \quad \approx \quad 
10^{-11} \left(  \frac{M}{M_P} \right)^3  \frac{ \left( \Lambda^3 \mu^3 - \mu^6 \right) }{ \Lambda^{3/2} \,  T^3_{\rm{RH}}  }.
\end{align}
Using the parametrization in Eq. \eqref{eq:parametrization}, we have
\begin{align}
\frac{m^3 \left( \delta^3 \Delta^3 \, 10^{12} - \Delta^6  \right) }{ \delta^{3/2} \gamma^3} \approx n \, 10^{22}
\end{align}
where $\mu = \Delta 10^8$~GeV, $\Delta \in \{ 1, 100  \}$ and $ \gamma \in \{ 10^{-2}, 1000  \} $. \ref{fig:reheatingGeneric} shows the behaviour of the reheating temperature as a function of $M$, for different ranges of $\Lambda, \mu$. From the plots we infer that the results obtained in Sec. \eqref{sec:baryon_asymm} by neglecting $\mu$ are consistent with the results 
in this Appendix.
\section{Effect of the Riemann coupling near Schwarzschild black hole }
\label{app:blackhole}

In this appendix, we will show that $S_{\rm CB}$ coupling is tiny near the
solar mass Schwarzschild black-holes. Specifically, we evaluate at the 
Schwarzschild radius of a non-rotating spherically symmetric black hole of mass $\mu$. Since the calculation is order of magnitude, we calculate the Kretschmann scalar $K = R_{\mu\nu\alpha\beta} R^{\mu\nu\alpha\beta}$ for the Schwarzschild black-hole. For this case, the Kretschmann scalar $K$ at a radial distance $r$ from the black-hole center is given by:
\begin{align}\label{Kretschmann-scalar}
K = \frac{48 G^2 \mu^2}{c^4 r^6}
\end{align}
which implies that the Riemann tensor $\sim \sqrt{K} $. The coupling term
\begin{align}
\frac{\sqrt{K}}{M^2} \sim \frac{\sqrt{48} G \mu}{c^2 r^3} \frac{1}{M^2}.
\end{align}
For Schwarzschild radius $r_h = {2G \mu}/{c^2}$, the coupling term becomes 
\begin{align}\label{app:Kretschmann-scalar-coupling}
\frac{\sqrt{K}}{M^2} \sim   \sqrt{\frac{3}{4}}  \frac{c^4}{ G^2 \, \mu^2 \, M^2 } \, .
\end{align}
We set $\mu = 1 M_\odot $ where $ M_\odot \approx 10^{30} \rm{Kg}$ is the solar mass. Since the result might be interesting to various astrophysical and cosmological phenomenon, we will calculate the value of the coupling $\sqrt{K}/M^2$ (\ref{app:Kretschmann-scalar-coupling}) in units which are preferred in early universe cosmology (natural units) and gravity (geometrized units).

Natural units are preferred in the early Universe Physics, and all scales are rewritten in terms of $GeV$. For simplicity, we write the following quantities in terms of $m_P$ as,
\begin{align}\label{app:NaturalUnits}
G \approx M_P^{-2}, \qquad c = 1, \qquad M \approx 10^{-2} M_P, \qquad \mu = 10^{30} \rm{Kg} \approx 10^{38} M_P 
\end{align}
where we have used $M_P \approx 10^{19} \rm{GeV} \approx 10^{-8} \rm{Kg}$ and $M = 10^{17} \rm{GeV} \approx 10^{-2} M_P$. Using Eq.(\ref{app:NaturalUnits}) in Eq.(\ref{app:Kretschmann-scalar-coupling}), we get 
\begin{align}\label{app:KS-coupling-NaturalUnits}
\frac{\sqrt{K}}{M^2} \approx 10^{-72}.
\end{align}
To understand the effect in astrophysical phenomenon, we calculate the value of the coupling (\ref{app:Kretschmann-scalar-coupling}) in geometrized units ($c = G = 1$). Since our model parameter $M \approx 10^{17} \rm{GeV}$ has unit of energy, we need to substitute the conversion $M \rightarrow \frac{M G}{c^4}$ in
Eq.~(\ref{app:Kretschmann-scalar-coupling}), which gives 
\begin{align}\label{app:KS-coupling-SI}
\frac{\sqrt{K}}{M^2} \approx   \sqrt{\frac{3}{4}}  \frac{c^{12} }{ G^4 \, \mu^2 \, M^2 }.
\end{align}
Using the following values,
\begin{align}
c = G = 1,  \qquad 
M = 10^{-10} \rm{Kg},  \qquad 
\mu \approx 10^{30} \rm{Kg}
\end{align}
we obtain
\begin{align}\label{app:KS-coupling-SI_Units}
\frac{\sqrt{K}}{M^2} \approx 10^{-40} \, \rm{Kg}^{-4}.
\end{align}
Hence, from the above analysis, it is clear that the effect near the solar mass size black-hole is negligible. Note that since Riemann tensor is coupled with gauge kinetic terms $F_{\mu\nu}F^{\mu\nu}$, the coupling term contains the frequency of the electromagnetic waves, i.e., $\omega^2$, hence the coupling will have effects at very high frequency. 
The coupling constant \eqref{app:Kretschmann-scalar-coupling} 
is of the order of one or greater for black-holes of mass $\mu \sim 100 M_P$, i.e.,

\begin{align}
 \frac{ \sqrt{K} }{M^2} \sim 1, \qquad \text{for} \quad \mu \sim 100 M_{P}  
\end{align}
Such size primordial black-holes form in the very early Universe. For such black-holes, Hawking temperature is~\cite{2008-Das.Shanki.Sur-Review} 
\begin{align}
T_H = \left(  \frac{\hbar c^3}{G k_B}  \right) \frac{1}{8\pi \mu} \quad \approx  \quad 10^{-3} M_P 
\end{align}
Since the Hawking radiation is thermal, we can obtain the peak wavelength of the black-body spectrum from Wien's displacement law $\lambda_{\rm{max} } T = \rm{constant}$. The wavelength $\lambda_{\rm{max} }$ corresponding to $100 M_P$ mass black holes is given by: 
\begin{align}
 \lambda_{\rm{max} } \sim T_H^{-1} \approx 10^3 M_P^{-1} 
\end{align}
Using the conversion $1~\rm{GeV} = 5.06 \times 10^{15} m^{-1}$ which gives $M_P \approx 10^{34} m^{-1}$, we get  
\begin{align}
 \lambda_{\rm{max} } \approx 10^3 M_P^{-1} \approx 10^{-31} m, \qquad \implies \nu = \frac{c}{\lambda_{\rm{max}} } = 10^{39} s^{-1}. 
\end{align}
Assuming these PBHs are produced just after bigbang, the above frequency will be redshifted by a factor $10^{20}$. Thus, the redshifted frequency is $\sim 10^{19} Hz$ (Gamma-ray) and can have potential signatures. However, it is important to note that at that scale, we need to include higher-order corrections. 
As mentioned earlier, in this analysis, we have ignored the mass dimension 8 operators~\cite{2019-Ruhdorfer.etal-JHEP}. At the Planck scale, we need to include dimension 8 and beyond. Hence, to understand these effects of PBH of Planck mass size, we need to include mass-dimension 8 operators and beyond.

\input{References.bbl}
\end{document}

%% file: References.bbl
%